\documentclass{article}

\usepackage{arxiv}

\usepackage[utf8]{inputenc} % allow utf-8 input
\usepackage[T1]{fontenc}    % use 8-bit T1 fonts
\usepackage{hyperref}       % hyperlinks
\usepackage{url}            % simple URL typesetting
\usepackage{booktabs}       % professional-quality tables
\usepackage{amsfonts}       % blackboard math symbols
\usepackage{nicefrac}       % compact symbols for 1/2, etc.
\usepackage{microtype}      % microtypography
\usepackage{lipsum}		% Can be removed after putting your text content
\usepackage{float}
\usepackage{amsmath}
\usepackage{graphicx}
\usepackage{doi}
\usepackage{subcaption}

\title{Oxygen, Angiogenesis, Cancer and Immune Interplay in Breast Tumor Micro-Environment: A Computational Investigation}

\author{Navid Mohammad Mirzaei \thanks{Corresponding author.} \\
	Department of Epidemiology\\
    Mailman School of Public Health\\
    Columbia University\\ New York, NY 10032, USA\\
	\texttt{nm3519@cumc.columbia.edu} \\
	\And
    Panayotis G. Kevrekidis \\
	Department of Mathematics and Statistics\\
	University of Massachusetts\\
	Amherst, MA 01003-4515, USA \\
	\texttt{kevrekid@umass.edu} \\
    \And
    Leili Shahriyari \thanks{Corresponding author.} \\
	Department of Mathematics and Statistics\\
	University of Massachusetts\\
	Amherst, MA 01003-4515, USA \\
	\texttt{lshahriyari@umass.edu} }
\renewcommand{\shorttitle}{}

\begin{document}
\maketitle

\begin{abstract}
Breast cancer is one of the most challenging global health problems among women. This study investigates the intricate breast tumor microenvironment (TME) dynamics utilizing data from Mammary-specific Polyomavirus Middle T Antigen Overexpression mouse models (MMTV-PyMT). It incorporates Endothelial Cells (ECs), oxygen, and Vascular Endothelial Growth Factors (VEGF) to examine the interplay of angiogenesis, hypoxia, VEGF, and the immune cells in cancer progression.
We introduce an approach to impute the immune cell fractions within the TME using single-cell RNA-sequencing (scRNA-seq) data from MMTV-PyMT mice. We further quantify our analysis by estimating cell counts using cell size data and laboratory findings from existing literature.
Parameter estimation is carried out via a Hybrid Genetic Algorithm (HGA).
Our simulations reveal various TME behaviors, emphasizing the critical role of adipocytes, angiogenesis, hypoxia, and oxygen transport in driving immune responses and cancer progression.
The global sensitivity analyses highlight potential therapeutic intervention points, such as VEGFs' critical role in EC growth and oxygen transportation and severe hypoxia's effect on the cancer and  the total number of cells. The VEGF-mediated production rate of ECs shows an essential time-dependent impact, highlighting the importance of early intervention in slowing cancer progression.  These findings align with the observations from the clinical trials demonstrating the efficacy of VEGF inhibitors and suggest a timely intervention for better outcomes.
\end{abstract}

% keywords can be removed

\section{Introduction}
Breast cancer is the most prevalent cancer among women globally \cite{kashyap2022global}, with 2.3 million new cases and 685,000 deaths reported in 2022 \cite{arnold2022current}. The estimated cases for the United States in 2023 are 298,000, with 43,000 deaths \cite{siegel2023cancer}. Surgical removal followed by radiotherapy is a standard treatment for non-metastatic breast cancer, varying based on subtypes such as HR+, ERBB2+, and triple-negative \cite{sharma2010various}. For example, endocrine therapy is prescribed for HR+ subtypes and trastuzumab-based ERBB2-directed antibody therapy in addition to chemotherapy for ERBB2+ subtypes and chemotherapy alone for triple-negative subtypes. 
Understanding the complex cellular and molecular interplays of the Tumor Micro-Environment (TME) is crucial in achieving more effective treatment strategies \cite{soysal2015role}. 

The TME, an ecosystem including tumor cells, immune cells, fibroblasts, blood vessels, cytokines, and the extracellular matrix, is an active promoter of cancer progression \cite{truffi2020fibroblasts}. 
New technologies are developed to study the intricacies within the TME, such as spatial transcriptomics and multiplexed proteomics \cite{elhanani2023spatial}. Longitudinal in-vivo investigations of the TME are costly and straining, so an alternative is using genetically engineered mouse models \cite{sharpless2006mighty,walrath2010genetically,kersten2017genetically}.

The mammary-specific polyomavirus middle T antigen overexpression mouse model (MMTV-PyMT) is one of the most popular mouse models since its discovery in 1992 \cite{attalla2021insights,browne2015runx1,hallett2013anti,tatarova2022multiplex}, developing spontaneous mammary tumors similar to human breast cancers \cite{christenson2017mmtv}. The histopathology and breast cancer biomarkers expression in the MMTV-PyMT tumors is in line with late-stage human breast cancers \cite{almholt2008metastasis}. 

Despite mouse models' convenience, making predictions about cells and cytokine dynamics in the TME is still challenging. 
However, mathematical models can help scientists make valuable estimations, predictions, or hypotheses based on limited data. Some mathematical cancer models focus on a certain cell type mutation \cite{makhlouf2020mathematical,mahlbacher2018mathematical}, some investigate a set of interactions in the TME \cite{mohammad2021mathematical,louzoun2014mathematical,Mirzaei2023},  others aim to discover optimal treatment dosage \cite{nani2000mathematical,budithi2021data, Le2021Investigating}, and some describe it as an epidemiological problem and search for the contributing risk factors \cite{brouwer2016age, li2023mathematical}.

Among the crucial mechanisms in TME benefiting from mathematical modeling are those leading to vessel formation (angiogenesis) and oxygen delivery. Chaplain has
introduced a generic tumor angiogenesis modelbased on the ECs, angiogenic cytokine, and fibronectin interactions \cite{chaplain2000mathematical}; see also the earlier impactful work of~\cite{andersonchaplain}. Harrington et al. designed a mathematical model to simulate vessel formation in the cornea in the presence of promoter and inhibitor cytokines \cite{harrington2007hybrid}, following the earlier work of~\cite{TONG200114}. Both these and other related (see, e.g., the review of~\cite{peirce}) studies focus on the Vessels' spatial distribution and obtain valuable results agreeing with relevant biological observations. Nevertheless, considerable space exists for further developments, notably the angiogenic factors (so-called
TAFs) and TME interactions.

Angiogenesis, a critical process in tumor progression, plays a central role in breast cancer development. Tumors require a constant blood supply to sustain their growth, and angiogenesis ensures new blood vessel formation. The solid TME often experiences hypoxia, a condition of insufficient oxygen levels. Hypoxia triggers the VEGF upregulation, a key pro-angiogenic factor. VEGF promotes EC proliferation and migration, leading to new blood vessel formation. The interactions in the TME can play a significant role in promoting or inhibiting angiogenesis. The intricate interplays between the TME, angiogenesis, hypoxia, and VEGF underscore the significance of targeting these pathways for therapeutic interventions. Numerous studies have investigated these mechanisms, with seminal works including Folkman \cite{folkman1971tumor} and Ferrara et al. \cite{ferrara2004discovery}. 

In this paper, we propose an ODE model describing the interactions in the TME, an extension of a previous study \cite{mohammad2022investigating}. The extension concerns angiogenesis and oxygen delivery mechanisms, i.e., we include three major players: ECs, oxygen, and VEGF. While spatial interactions among these elements are not considered for simplicity, we acknowledge their significance as a potential future research
direction.
Here, we present a methodology to prepare scRNA-seq data for an ODE system describing TME interactions. Using HGA, we estimate parameters and discuss simulation outcomes, augmented by sensitivity analyses to uncover influential mechanisms on the system, cancer cells, and total immune cells. Our findings highlight the significance of pathways related to angiogenesis, oxygenation, and hypoxia.

\subsection{The model}
\begin{figure}[H]
    \centering
    \includegraphics[scale=0.39]{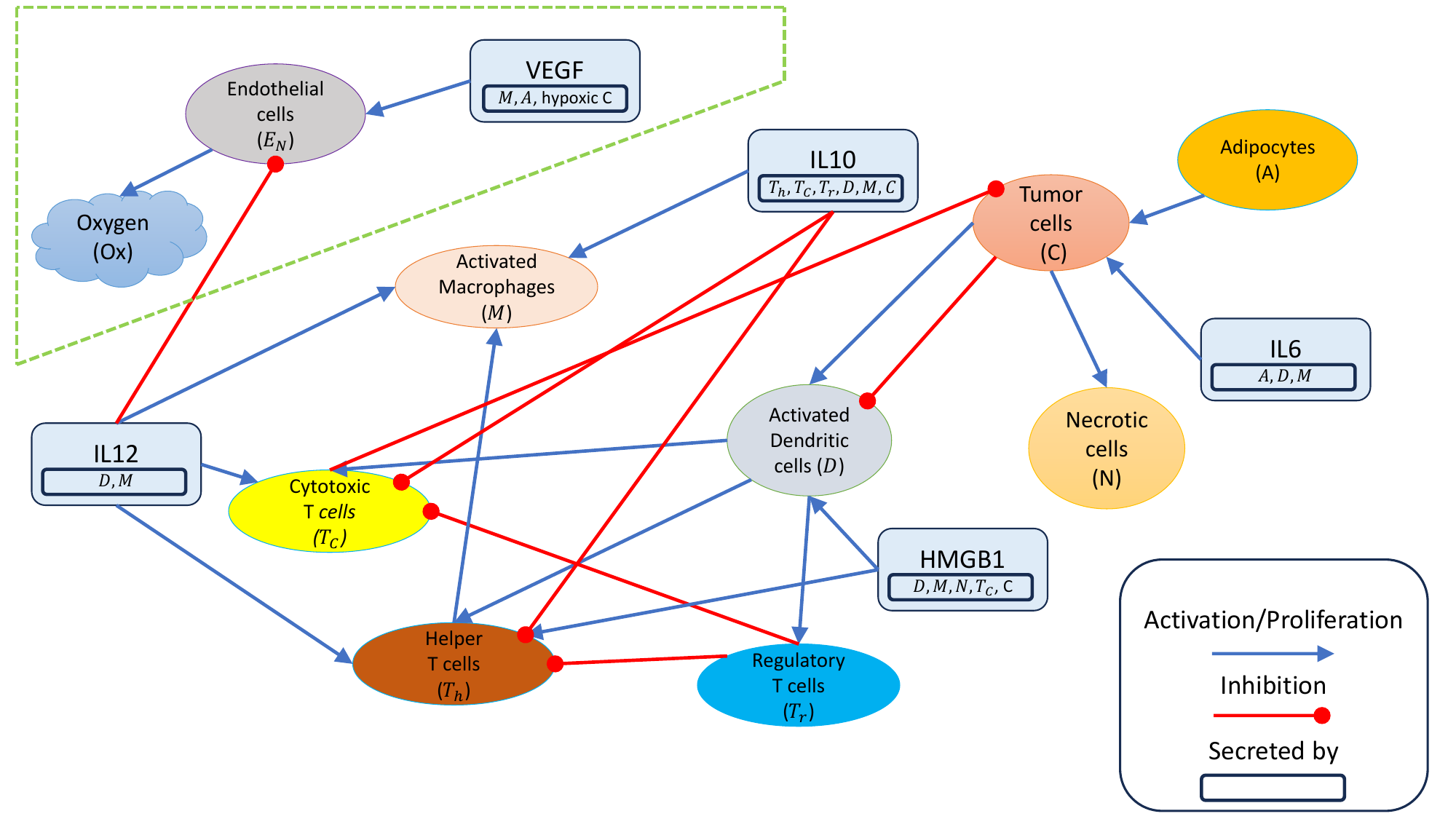}
    \caption{Interaction network. The portion inside the dashed shape is an addition to the network used by Mohammad Mirzaei et al. \cite{mohammad2022investigating, mohammad2022pde}. Despite the importance of oxygen in the network, no arrows are coming out of it. This is because lack of oxygen (hypoxia) is the key factor, not oxygen itself.}
    \label{fig:network}
\end{figure}

The model in this paper extends the framework established by Mohammad Mirzaei et al. \cite{mohammad2022investigating}. In our adaptation, we have incorporated additional state variables: VEGF, ECs, and Oxygen. These new elements are added to investigate hypoxia and angiogenesis roles in tumor progression. For a detailed representation of the interaction network, see Figure~\ref{fig:network}. The newly incorporated interactions are distinctly marked for easy identification. Table \ref{table:var} shows the model variables with their corresponding equations.

The naming conventions for our parameters are as follows:
\begin{eqnarray*}
    &\lambda_{XY}& \quad \text{The rate molecule/cell Y promotes the} \\ & & \quad  \text{production or proliferation of molecule/cell X} \\
    &\delta_{XY}& \quad \text{The rate molecule/cell Y inhibits the } \\ & & \quad  \text{production or proliferation of molecule/cell X} \\
    &\delta_{XYOx}& \quad \text{The rate molecule/cell Y inhibits the} \\ & & \quad  \text{production of molecule/cell X proportional to} \\ & & \quad  \text{the hypoxia function value} \\
    &\delta_X& \quad \text{Natural decay/death rate of molecule/cell X} \\
    &A_X& \quad \text{Innate production rate of the cell X} \\
    &\delta_i& \quad \text{For } i=1,\cdots,6. \text{ Oxygen consumption rate by} \\ & & \quad  \text{the six major cell types in the model}
\end{eqnarray*}
In this paper, parameters with overhead bars are in non-dimensionalized form. Mathematically, we consider production processes, both for a single species or stimulated by the presence of a second species. Additionally, we account for degradation processes, either occurring naturally for a single species or stimulated by the presence of or interaction with another species.

\begin{figure}[H]
    \centering
    \includegraphics[scale=0.45]{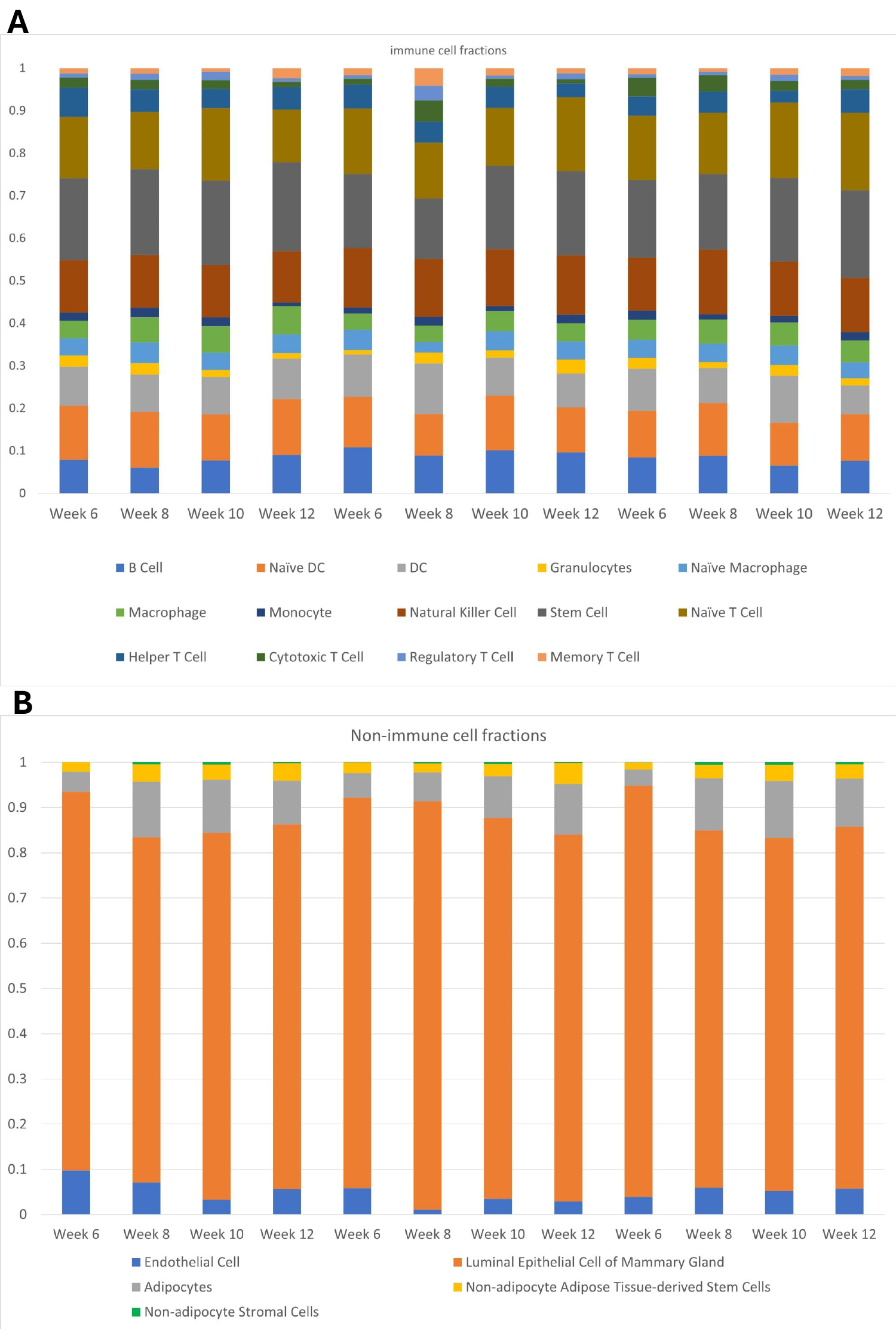}
    \caption{Cell fractions for (A)Immune cells and (B) Non-immune cells. The proportions are calculated for three mice at week 6 (hyperplasia), week 8 (adenoma/MIN), week 10 (early carcinoma), and week 12 (late carcinoma).}
    \label{fig:cell-fracs}
\end{figure}

We used RNA-sequencing data from three MMTV-PyMT mice \cite{cai2017transcriptomic}. The proportions of various immune cells were estimated using the digital cytometry technique CIBERSORTx \cite{newman2019determining}, with the signature matrix extracted from the Immunological Genome Project \cite{heng2008immunological}. We created a unique signature matrix for non-immune cell populations using single-cell RNA sequencing data from the Tabula Muris database \cite{schaum2018single}. Figure \ref{fig:cell-fracs} show imputed fractions for immune and non-immune cells. Additionally, we applied a laboratory protocol tailored for MMTV-PyMT mice to quantify macrophages and cancer cells relative to tumor size. This process approximated the total cell count, informed by imputed cell fractions, and allowed the direct extraction of cytokine levels from the RNA-sequencing data.

\begin{table}[H]
\centering
	\resizebox{0.95\linewidth}{!}{
		\begin{tabular}{|c|l|l|}
			\hline 
			\textbf{Variable} & \textbf{Name} & \textbf{Equation for their rate of changes} \tabularnewline
			\hline 
			\hline 
            $E_N$ & ECs & $\begin{array}{l} 
            \\\frac{d[\overline{E_N}]}{dt} = \overline{\lambda}_{VE_N}[\overline{V}]\Big(1-\frac{[\overline{E_N}]}{\overline{E_N}_0}\Big)[\overline{E_N]}-\overline{\delta}_{E_N IL_{12}}[\overline{IL_{12}}][\overline{E_N}]\\
            -\Big({\delta}_{E_N}+\overline{\delta}_{E_N O_x} f_2([\overline{O_x}])\Big)[\overline{E_N}]\vspace{0.1in}
            \end{array}$\\
			\hline
            $V$ & VEGF & $\begin{array}{l} \\
            {{\frac{d[\overline{V}]}{dt}}} = {{\overline{\lambda}_{V A}}  [\overline{A}] + {\overline{\lambda}_{V M}}  [\overline{M}] +  {{\overline{\lambda}_{VCO_x}[\overline{C}]f_1(O_x) }}+{{\overline{\lambda}_{VO_x} f_1([\overline{O_x}])}}} - {\delta_{V} [\overline{V}]}\vspace{0.1in}
            \end{array}$\\
			\hline
            $O_x$ & Oxygen & $\begin{array}{l}\\
            {{\frac{d[\overline{O_x}]}{d\tau}}} = \overline{\lambda}_{O_xE_N}([\overline{E_N}]) - \overline{\delta}_1 \left([\overline{T_N}]+[\overline{T_h}]+[\overline{T_c}]+ [\overline{T_r}]\right)[\overline{O_x}]-\overline{\delta}_2 \left([\overline{D_N}]+ [\overline{D}] \right)[\overline{O_x}]\\-\overline{\delta}_3\left( [\overline{M_N}]+[\overline{M}]\right)[\overline{O_x}]-\overline{\delta}_{4} [\overline{C}][\overline{O_x}]-\overline{\delta}_{5} [\overline{A}][\overline{O_x}]-\overline{\delta}_{6} [\overline{E_N}][\overline{O_x}]-\delta_{O_x}[\overline{O_x}]\vspace{0.1in}
            \end{array}$\\
			\hline
			$T_{N}$ & Naive T-cells & 
            $\begin{array}{l}\\
    \frac{d [\overline{T_N}]}{d t} =
\overline{A}_{T_N} - 
\left( \overline{\lambda}_{T_h H}[\overline{H}]
+ \overline{\lambda}_{T_h D} [\overline{D}]
+ \overline{\lambda}_{T_h IL_{12}} [\overline{IL_{12}}]
\right)[\overline{T_N}]
-\left(\overline{\lambda}_{T_c D}[\overline{D}] 
+ \overline{\lambda}_{T_c IL_{12}}[\overline{IL_{12}}]
\right)[\overline{T_N}]\\-\left( \overline{\lambda}_{T_r D}[\overline{D}] + \overline{\lambda}_{T_r Ox} f_1([\overline{Ox}])\right)[\overline{T_N}]-\left(\delta_{T_N}+\overline{\delta}_{T_NOx}f_2([\overline{Ox}])\right)[\overline{T_N}]\vspace{0.1in}
\end{array}$\\
			\hline 
			$T_{h}$ & Helper T-cells &  $\begin{array}{l} \\
            \frac{d [\overline{T_h}]}{d t} =
\left(\overline{\lambda}_{T_h H}[\overline{H}] 
+ \overline{\lambda}_{T_h D} [\overline{D}] 
+ \overline{\lambda}_{T_h IL_{12}} [\overline{IL_{12}}] 
\right) [\overline{T_N}] - \left(\overline{\delta}_{T_h T_r} [\overline{T_r}]
+ \overline{\delta}_{T_h IL_{10}} [\overline{IL_{10}}]
\right)[\overline{T_h}]\\-\left({{\delta}_{T_h}}+{{\overline{\delta}_{T_h O_x} f_2([\overline{O_x}])}}\right)[\overline{T_h}]\vspace{0.1in}
            \end{array}$\\
			\hline 
			$T_{C}$ & Cytotoxic cells & $\begin{array}{l} \\
            \frac{d[\overline{T_c}]}{dt} = 
\left(\overline{\lambda}_{T_c D}[\overline{D}] 
+  \overline{\lambda}_{T_c IL_{12}}[\overline{IL_{12}}]
\right)[\overline{T_N}]
-\left( 
\overline{\delta}_{T_c T_r}[\overline{T_r}] + \overline{\delta}_{T_c IL_{10}} [\overline{IL_{10}}]\right)[\overline{T_c}]\\-\left({{\delta}_{T_c}}+{{\overline{\delta}_{T_c O_x} f_2([\overline{O_x}])}}\right)[\overline{T_c}]\vspace{0.1in}
            \end{array}$\\
			\hline 
			$T_{r}$ & Regulatory T-cells & $\begin{array}{l}\\ 
            \frac{d[\overline{T_r}]}{dt} = \left(
    \overline{\lambda}_{T_r D}[\overline{D}]
     + \overline{\lambda}_{T_r O_x} \ f_1([\overline{O_x}]) \right)\ [\overline{T_N}] - \left({{\delta}_{T_r}}+{{\overline{\delta}_{T_r O_x} f_2([\overline{O_x}])}}\right)[\overline{T_r}]\vspace{0.1in}
            \end{array}$\\
			\hline 
			$D_{N}$ & Naive DCs &$\begin{array}{l} \\
            \frac{d [\overline{D_N}]}{dt} =
    \overline{A}_{D_N} 
    - \left(\overline{\lambda}_{D C}[\overline{C}] 
    + \overline{\lambda}_{D H}[\overline{H}]+ {{\overline{\lambda}_{DO_x} f_{1}([\overline{O_x}])}}\right)[\overline{D_N}] 
    - \left({{\delta}_{D_N}}+{{\overline{\delta}_{D_N O_x} f_{2}([\overline{O_x}])}}\right)[\overline{D_N}]\vspace{0.1in}
            \end{array}$\\
			\hline 
			$D$ & Activated DCs &  $\begin{array}{l} \\
             \frac{d [\overline{D}]}{d t} = 
    \left(\overline{\lambda}_{D C}[\overline{C}] 
    + \overline{\lambda}_{D H}[\overline{H}] + {{\overline{\lambda}_{DO_x} f_{1}([\overline{O_x}])}}
    \right)[\overline{D_N}]- \left(\overline{\delta}_{D C}[\overline{C}] 
    + {{\delta}_{D}}+{{\overline{\delta}_{D_N O_x} f_{2}([\overline{O_x}])}}\right)[\overline{D}]\vspace{0.1in}
            \end{array}$\\
			\hline 
			$M_{N}$ & Naive Macrophages & $\begin{array}{l} \\
            \frac{d [\overline{M_N}]}{d t} =
\overline{A}_{M_N} - 
\Big(
\overline{\lambda}_{M IL_{10}} [\overline{IL_{10}}]
+ \overline{\lambda}_{M IL_{12}} [\overline{IL_{12}}]+\overline{\lambda}_{M T_h} [\overline{T_h}]
+{\delta}_{M_N}\Big) [\overline{M_N}]\vspace{0.1in}
            \end{array}$\\
			\hline 
			$M$ & Macrophages & $\begin{array}{l} \\
            \frac{d [\overline{M}]}{d t} =
\left(
\overline{\lambda}_{M IL_{10}} [\overline{IL_{10}}]
+ \overline{\lambda}_{M IL_{12}} [\overline{IL_{12}}]
+ \overline{\lambda}_{M T_h} [\overline{T_h}]
\right) [\overline{M_N}]-
{{\delta}_{M}}[\overline{M}]\vspace{0.1in}
            \end{array}$\\
			\hline 
			$C$ & Cancer cells & $\begin{array}{l} \\
            \frac{d[\overline{C}]}{d t}  = \left(
    {\lambda}_C  + 
    \overline{\lambda}_{C IL_6}[\overline{IL_6}] +
    \overline{\lambda}_{C A}[\overline{A}]
    \right) \left(1- \frac{[\overline{C}]}{\overline{C_0}}\right)[\overline{C}] -\left(\overline{\delta}_{CT_c}[\overline{T_c}]+{{\overline{\delta}_{C T_c O_x} f_1([\overline{O_x}])}}[\overline{T_c}]\right) [\overline{C}]\\-\left(    {{\overline{\delta}_{CO_x}f_2([\overline{O_x}])}}+\delta_C\right) [\overline{C}]\vspace{0.1in}
            \end{array}$\\
			\hline 
			$N$ & Necrotic cells & $\begin{array}{l} \\
            \frac{d[\overline{N}]}{dt}  = \overline{\alpha}_{N C}
    \left(\overline{\delta}_{CT_c}[\overline{T_c}]+{{\overline{\delta}_{C T_c O_x} f_1([\overline{O_x}])}}[\overline{T_c}] +{{\overline{\delta}_{CO_x}f_2([\overline{O_x}])}}+\delta_C\right) [\overline{C}]
    -{\delta}_N[\overline{N}]\vspace{0.1in}
            \end{array}$\\
			\hline 
			$A$ & Adipocytes & $\begin{array}{l} \\
            \frac{d[\overline{A}]}{dt}  = {\lambda}_A\left(1-\frac{[\overline{A}]}{\overline{A_0}}\right) [\overline{A}]- \delta_A[\overline{A}]\vspace{0.1in}
            \end{array}$\\
			\hline 
			$H$ & HMGB1 &  $\begin{array}{l} \\
            \frac{d[\overline{H}]}{dt}  = \overline{\lambda}_{HD}[\overline{D}] + \overline{\lambda}_{HN}[\overline{N}] + \overline{\lambda}_{HM}[\overline{M}] + \overline{\lambda}_{HT_c}[\overline{T_c}] + \overline{\lambda}_{HC}[\overline{C}] - {\delta}_H[\overline{H}]\vspace{0.1in}
            \end{array}$\\
			\hline
			$IL_{12}$ & IL-12 & $\begin{array}{l} \\
            \frac{d [\overline{IL_{12}}]}{d t}  = 
\overline{\lambda}_{IL_{12} M}[\overline{M}] 
+ \overline{\lambda}_{IL_{12} D} [\overline{D}]  
% + \overline{\lambda}_{IL_{12} T_h} [\overline{T_h}]  
% + \overline{\lambda}_{IL_{12} T_c} [\overline{T_c}]
-  {\delta}_{IL_{12}} [\overline{IL_{12}}]\vspace{0.1in}
            \end{array}$\\
			\hline 
			$IL_{10}$ & IL-10 & $\begin{array}{l} \\
            \frac{d [\overline{IL_{10}}]}{d t}  = 
\overline{\lambda}_{IL_{10} M}[\overline{M}]    
+ \overline{\lambda}_{IL_{10} D} [\overline{D}]  
+ \overline{\lambda}_{IL_{10} T_r} [\overline{T_r}] 
+ \overline{\lambda}_{IL_{10} T_h} [\overline{T_h}]  
+ \overline{\lambda}_{IL_{10} T_c} [\overline{T_c}]
 + \overline{\lambda}_{IL_{10} C} [\overline{C}] \\+ {{\overline{\lambda}_{IL_{10} O_x}  f_1([\overline{O_x}])}} 
- {\delta}_{IL_{10}} [\overline{IL_{10}}]\vspace{0.1in}
            \end{array}$\\
			\hline
			$IL_{6}$ & IL-6 & $\begin{array}{l} \\
            \frac{d[\overline{IL_6}]}{d t}  = 
\overline{\lambda}_{IL_6 A}  [\overline{A}]  
+ \overline{\lambda}_{IL_6 M} [\overline{M}] +{{\overline{\lambda}_{IL_6 M O_x} [\overline{M}]f_1([\overline{O_x}])  }}
+ \overline{\lambda}_{IL_6 D}  [\overline{D}] + {{\overline{\lambda}_{IL_6 O_x}  f_1([\overline{O_x}])}}
\\- \delta_{IL_6} [\overline{IL_6}]\vspace{0.1in}
            \end{array}$\\
			\hline
\end{tabular}}
 \caption{{ Model variables with their equations.} }
 \label{table:var}
\end{table}

We employed a Hybrid Genetic Algorithm (HGA) for parameter estimation in our model, followed by a global sensitivity analysis to determine parameters significantly affecting overall system dynamics, cancer and total immune cell counts. Perturbing these sensitive parameters allows us to assess and quantify their impact on model outcomes, as detailed in the Methods section.

\section{Methods}
\subsection*{Model variables and their interactions}
% We extend the breast TME model provided in \cite{mohammad2022investigating} by adding hypoxia and angiogenesis-related components, including ECs, VEGF, and Oxygen. 
We use the mass action law to create the ODE system for the interactions among variables as discussed in \cite{mirzaei2023modeling}.

\subsubsection*{Hypoxia}
Hypoxic effects take hours to impact the system~\cite{liu2022acute}, while the simulation time for our problem and the associated data is on the order of weeks. Therefore, we assume low oxygen levels entail hypoxic effects on cells for {each computational time step corresponding to one day. }
%But what counts as a low oxygen level?
Usually, laboratory cell cultures are in a hyperoxic condition with about 20.9\% oxygenation \cite{muz2015role}. However, the in-vivo normoxic oxygen level in breast tissue is about 8.5\%, and the hypoxic level is below 1\% \cite{vaupel2007detection}. Although different T-cell types have different hypoxic responses at different oxygen levels, most of these hypoxic activities occur in the range 0\%-1\% \cite{ebbesen2004linking}. In particular, at 0.1\%-1\%, there is not enough oxygen in the environment to disintegrate HIF-1$\alpha$, so the latter will accumulate and trigger some hypoxic functions and below 0.1\% most cell cycles yield, causing hypoxic death \cite{ebbesen2004linking}. For simplicity, if a particular cell or cytokine is promoted or inhibited via HIF pathways, which are not included in our model, we use a rate independent of other state variables and only dependent on oxygen levels. Some hypoxia outcomes through HIF pathways included in our model are up-regulating the VEGF, IL-6, and IL-10 production \cite{noman2015hypoxia}, promoting the maturation of the dendritic cells \cite{liu2019ccr7,filippi2014short}, increasing the regulatory T-cells differentiation \cite{chen2022role,clambey2012hypoxia} and increasing the T-cells cytotoxicity \cite{emami2021role,gropper2017culturing}. These outcomes indirectly lead to reduced cytotoxic and helper T-cell differentiation. Also, hypoxia promotes the macrophages' differentiation into a pro-tumor subtype \cite{colegio2014functional}. However, since we do not include the macrophage subtypes in our model, we include this effect through the IL-6 pathway. In other words, we assume hypoxia promotes IL-6 production by macrophages, which has a strong pro-inflammatory effect. 

In this paper, we assume blood vessel formation is proportional to the number of endothelial cells (ECs), facilitating oxygen transportation. Thus, oxygen serves as a source dependent on ECs. Regions with oxygen levels below a critical threshold are considered hypoxic. We specify two hypoxic states: mild hypoxia initiating the HIF pathways (between 0.1\%-1\% oxygen) and severe hypoxia causing cell cycle arrest (below 0.1\%). Cells can die through mechanisms in the original model (via inhibitors or normoxic death) or undergo hypoxic death. To capture the promotion or inhibition effects of hypoxia, we define the following function:
\begin{equation}
  f_i(x) =   \frac{(Ox_{crit}^{(i)})^m}{(Ox_{crit}^{(i)})^m+x^m}, \qquad  i=1,2\label{eq:hypoxia-func}
\end{equation}
where $Ox_{crit}^{(i)}$ are the critical oxygen values and $m$ is the Hill-like coefficient, which governs the decrease rate from the maximum to the minimum function value. Notice that two functions are defined in \eqref{eq:hypoxia-func}, corresponding to two critical oxygen values. The first critical value corresponds to the transitioning threshold from normoxia to mild hypoxia, and the second one marks the transition from mild to severe hypoxia.
\subsubsection*{T-cells}
We classify T-cells into naive ($T_N$), helper ($T_h$), cytotoxic ($T_C$), and regulatory ($T_r$) types. Activated subtypes differentiate from the naive subtype, introducing negative feedback terms proportional to their level and corresponding activator levels. These terms serve as positive feedback for the activated subtypes, indicating regulation by the available number of naive subtypes. We employ a similar method for naive and activated dendritic cells and macrophages.

We consider a constant intrinsic proliferation rate for naive T-cells. Naive T-cells diminish due to differentiation or normoxic and hypoxic cell death. 

Helper T-cells' differentiation from naive T-cells is promoted by HMGB1 \cite{dong2022targeting,zhang2011potential}, dendritic cells \cite{steinman2007taking, hansen2017role}, and IL-12 \cite{knutson200412, cirella2022novel}. They are inhibited by regulatory T-cells \cite{beyer2006regulatory,whiteside2012regulatory} and IL10 \cite{salkeni2023interleukin,sheikhpour2018survey}. We also consider normoxic and hypoxic cell death for helper T-cells. 

Similarly, cytotoxic T-cells are promoted by dendritic cells \cite{fuertes2011host,lee2019role} and IL-12 \cite{cirella2022novel,mirlekar202112}. They are inhibited by regulatory T-cells \cite{watanabe2010regulatory,beyer2006regulatory} and IL-10 \cite{mirlekar202112,sato2011interleukin}. We consider normoxic and hypoxic cell death for these T-cells, too. 

Like the other types, regulatory T-cell differentiation is promoted by dendritic cells \cite{palucka2013dendritic,tran2015tumor}. Also, is mentioned in the ``Hypoxia" subsection, mild hypoxia can increase regulatory T-cell differentiation. Similarly to the
case of the other subtypes, we close the ODE by the normoxic and hypoxic cell death. 

Note, the hypoxic cell death is modeled via $f_2$, but the hypoxic promotion of $T_r$ is modeled via $f_1$, corresponding to severe and mild hypoxia, respectively. The corresponding ODEs in this case read:
\begin{align}
\frac{d [\overline{T_N}]}{d t}& =
\overline{A}_{T_N}  \nonumber \\
&- \left( \overline{\lambda}_{T_h H}[\overline{H}]
+ \overline{\lambda}_{T_h D} [\overline{D}]
+ \overline{\lambda}_{T_h IL_{12}} [\overline{IL_{12}}]
\right)[\overline{T_N}] \nonumber \\
&-\left(\overline{\lambda}_{T_c D}[\overline{D}] 
+ \overline{\lambda}_{T_c IL_{12}}[\overline{IL_{12}}]
\right)[\overline{T_N}] \nonumber \\
&-\left( \overline{\lambda}_{T_r D}[\overline{D}] + \overline{\lambda}_{T_r Ox} f_1([\overline{Ox}])\right)[\overline{T_N}]\nonumber \\
&-\left(\delta_{T_N}+\overline{\delta}_{T_NOx}f_2([\overline{Ox}])\right)[\overline{T_N}] \label{eq:TN}\\
\frac{d [\overline{T_h}]}{d t}& =
X\left(\overline{\lambda}_{T_h H}[\overline{H}] 
+ \overline{\lambda}_{T_h D} [\overline{D}] 
+ \overline{\lambda}_{T_h IL_{12}} [\overline{IL_{12}}] 
\right) [\overline{T_N}] \nonumber\\
&- \left(\overline{\delta}_{T_h T_r} [\overline{T_r}]
+ \overline{\delta}_{T_h IL_{10}} [\overline{IL_{10}}]
\right)[\overline{T_h}]\nonumber \\
&-\left({{\delta}_{T_h}}+{{\overline{\delta}_{T_h O_x} f_2([\overline{O_x}])}}\right)[\overline{T_h}],\label{eq:Th}\\
\frac{d[\overline{T_c}]}{dt} &= 
\left(\overline{\lambda}_{T_c D}[\overline{D}] 
+  \overline{\lambda}_{T_c IL_{12}}[\overline{IL_{12}}]
\right)[\overline{T_N}] \nonumber \\
&-\left( 
\overline{\delta}_{T_c T_r}[\overline{T_r}] + \overline{\delta}_{T_c IL_{10}} [\overline{IL_{10}}]\right)[\overline{T_c}]\nonumber \\
&-\left({{\delta}_{T_c}}+{{\overline{\delta}_{T_c O_x} f_2([\overline{O_x}])}}\right)[\overline{T_c}],\label{eq:Tc}
\\
\frac{d[\overline{T_r}]}{dt} &= \left(
    \overline{\lambda}_{T_r D}[\overline{D}]
     + \overline{\lambda}_{T_r O_x} \ f_1([\overline{O_x}]) \right)\ [\overline{T_N}] \nonumber \\
     &- \left({{\delta}_{T_r}}+{{\overline{\delta}_{T_r O_x} f_2([\overline{O_x}])}}\right)[\overline{T_r}],
\label{eq:Tr}
\end{align}
\subsubsection*{Dendritic cells}
We model two dendritic cell types: naive  ($D_N$) and activated ($D$) dendritic cells. Like T-cells, the activated dendritic cells differentiate from the naive ones. We also include normoxic and hypoxic cell death for both subtypes. Like naive T-cells, we consider a constant intrinsic growth rate for the naive dendritic cells. Their number decreases as they get activated or die. The tumor burden produces danger signals that promote dendritic cell maturation \cite{marciscano2021role}. HMGB1 is another factor contributing to dendritic activation \cite{tripathi2019hmgb1,campana2008hmgb1} and so is hypoxia, as explained in the "Hypoxia" subsection. Despite increasing the dendritic cell maturation, cancer cells can release factors inhibiting dendritic cell maturation and function \cite{tran2015tumor}.  
\begin{align}
        \frac{d [\overline{D_N}]}{dt} &=
    \overline{A}_{D_N} \nonumber \\
    &- \left(\overline{\lambda}_{D C}[\overline{C}] 
    + \overline{\lambda}_{D H}[\overline{H}]+ {{\overline{\lambda}_{DO_x} f_{1}([\overline{O_x}])}}\right)[\overline{D_N}] \nonumber \\
    &- \left({{\delta}_{D_N}}+{{\overline{\delta}_{D_N O_x} f_{2}([\overline{O_x}])}}\right)[\overline{D_N}], \label{eq:DN}
\\
    \frac{d [\overline{D}]}{d t} &= 
    \left(\overline{\lambda}_{D C}[\overline{C}] 
    + \overline{\lambda}_{D H}[\overline{H}] + {{\overline{\lambda}_{DO_x} f_{1}([\overline{O_x}])}}
    \right)[\overline{D_N}]\nonumber \\&- \left(\overline{\delta}_{D C}[\overline{C}] 
    + {{\delta}_{D}}+{{\overline{\delta}_{D_N O_x} f_{2}([\overline{O_x}])}}\right)[\overline{D}],
\label{eq:D}
\end{align}
\subsubsection*{Macrophages}
We consider naive ($M_N$) and activated macrophages ($M$) in the model. For simplicity, the different activated macrophages, such as M1 and M2 types, are consolidated as $M$. 
Differentiation of these types along the lines (e.g.,
see~\cite{bullbyrne}) would be an interesting
topic for further study.
Furthermore, we do not include hypoxic cell death for macrophages. Since these cell types tend to accumulate in hypoxic regions, they quickly adapt to severe hypoxia \cite{hao2012macrophages}. Like the other naive cell types, we model naive macrophages to proliferate at an innate constant rate. Their activation is augmented by IL-10 \cite{mirlekar2022tumor,muller2022mouse}, IL-12 \cite{mirlekar202112,steding2011role} and helper T-cells \cite{aras2017tameless}. Both macrophage subtypes die via normoxic cell death. Activated macrophage polarity is defined through its interaction with cancer cells, which is why they are often referred to as anti (M1) and pro (M2) tumor macrophages \cite{chen2019tumor}. However, as mentioned above, we do not explicitly include these subtypes in our model, and their different interaction with tumor cells is modeled through their cytokine pathways.

\begin{align}
    \frac{d [\overline{M_N}]}{d t} &=
\overline{A}_{M_N}- \Big(\overline{\lambda}_{M IL_{10}} [\overline{IL_{10}}]
+ \overline{\lambda}_{M IL_{12}} [\overline{IL_{12}}] \nonumber \\
&+\overline{\lambda}_{M T_h} [\overline{T_h}]
+{\delta}_{M_N}\Big) [\overline{M_N}], \label{eq:MN}
\\
\frac{d [\overline{M}]}{d t} &=
\Big(
\overline{\lambda}_{M IL_{10}} [\overline{IL_{10}}]
+ \overline{\lambda}_{M IL_{12}} [\overline{IL_{12}}] \nonumber \\
&+ \overline{\lambda}_{M T_h} [\overline{T_h}]
\Big) [\overline{M_N}]-
{{\delta}_{M}}[\overline{M}],
\label{eq:M}  
\end{align}
\subsubsection*{Cancer and Necrotic cells}
Cancer cells (C) grow according to a logistic model with carrying capacity $C_0$. Many factors can promote cancer proliferation, among which we have explicitly included the IL-6 \cite{di20146,guo2012interleukin} and adipocyte effects \cite{wang2019exosomes, gao2020adipocytes}. The other factors are considered via a constant rate $\lambda_C$. Cancer cells can be cleared via cytotoxic cells \cite{durgeau2018recent,raskov2021cytotoxic}. Also, as mentioned, T-cell cytotoxicity increases through HIF pathways, which is modeled by the term including $f_1([\overline{O_x}])$. Finally,  we include hypoxic and normoxic cell death mechanisms. Even though cancer cells proliferate much faster than normal cells, factors such as physical space, fitness landscapes, and nutrition impose a carrying capacity on their growth \cite{west2019cellular, rozhok2019generalized}. We apply the same rationale to the other two cell types (described in the following sections), namely adipocytes and ECs.

A portion of cancer cell death turns into necrotic cells \cite{liu2020necroptosis}. That portion is modeled via the constant $\overline{\alpha}_{N C}$. We only consider regular cell clearance for necrotic cells since they are dead cells and do not go through hypoxia.
\begin{align}
    \frac{d[\overline{C}]}{d t}  &= \left(
    {\lambda}_C  + 
    \overline{\lambda}_{C IL_6}[\overline{IL_6}] +
    \overline{\lambda}_{C A}[\overline{A}]
    \right) \left(1- \frac{[\overline{C}]}{\overline{C_0}}\right)[\overline{C}] \nonumber \\
    &-\left(\overline{\delta}_{CT_c}[\overline{T_c}]+{{\overline{\delta}_{C T_c O_x} f_1([\overline{O_x}])}}[\overline{T_c}]\right) [\overline{C}]\nonumber \\
    &-\left({{\overline{\delta}_{CO_x}f_2([\overline{O_x}])}}+\delta_C\right) [\overline{C}],
\label{eq:C}
\\
   \frac{d[\overline{N}]}{dt}  &= \overline{\alpha}_{N C}
    \Big(\overline{\delta}_{CT_c}[\overline{T_c}]+{{\overline{\delta}_{C T_c O_x} f_1([\overline{O_x}])}}[\overline{T_c}] \nonumber \\
    &+{{\overline{\delta}_{CO_x}f_2([\overline{O_x}])}}+\delta_C \Big) [\overline{C}]
    -{\delta}_N[\overline{N}], \label{eq:N}
\end{align}
\subsubsection*{Adipocytes}
Adipocytes contribute to cancer progression through intricate mechanisms. Having that contribution included in \eqref{eq:C} as $\overline{\lambda}_{C A}[\overline{A}]$, for simplicity, we model the adipocytes dynamics through a simple logistic model. 
\begin{align}
    \frac{d[\overline{A}]}{dt}  &= {\lambda}_A\left(1-\frac{[\overline{A}]}{\overline{A_0}}\right) [\overline{A}]- \delta_A[\overline{A}],
\label{eq:A}
\end{align}

\subsubsection*{Cytokines}
HMGB1 is produced by dendritic cells, necrotic cells, macrophages, cytotoxic T-cells, and cancer cells \cite{campana2008hmgb1,kang2013hmgb1}. IL-12 is secreted by dendritic cells and macrophages \cite{ma2015regulation}. IL-10 is a product of dendritic cells, macrophages, cancer cells, and cytotoxic, helper, and regulatory T-cells \cite{ma2015regulation}. Finally, adipocytes, dendritic cells, and macrophages are the main  IL-6 producers \cite{chen20226,luan2023adipocyte}. Note we have a hypoxic promotion of IL-6 production by macrophages. As mentioned in the "Hypoxia" section, this is to model the increase in pro-tumor macrophage differentiation via mild hypoxia. Additionally, IL-10 and IL-6 increase through HIF pathways, as explained in the "Hypoxia" section. We include a natural decay term for cytokine clearance.
\begin{align}
    \frac{d[\overline{H}]}{dt}  &= \overline{\lambda}_{HD}[\overline{D}] + \overline{\lambda}_{HN}[\overline{N}] + \overline{\lambda}_{HM}[\overline{M}] + \overline{\lambda}_{HT_c}[\overline{T_c}] \nonumber\\ &+ \overline{\lambda}_{HC}[\overline{C}] - {\delta}_H[\overline{H}],\label{eq:H}
\\
\frac{d [\overline{IL_{12}}]}{d t}  &= 
\overline{\lambda}_{IL_{12} M}[\overline{M}] 
+ \overline{\lambda}_{IL_{12} D} [\overline{D}]  
% + \overline{\lambda}_{IL_{12} T_h} [\overline{T_h}]  
% + \overline{\lambda}_{IL_{12} T_c} [\overline{T_c}]
-  {\delta}_{IL_{12}} [\overline{IL_{12}}],
\label{eq:IL-12}
\\
\frac{d [\overline{IL_{10}}]}{d t}  &= 
\overline{\lambda}_{IL_{10} M}[\overline{M}]    
+ \overline{\lambda}_{IL_{10} D} [\overline{D}]  
+ \overline{\lambda}_{IL_{10} T_r} [\overline{T_r}] \nonumber\\
&+ \overline{\lambda}_{IL_{10} T_h} [\overline{T_h}]  
+ \overline{\lambda}_{IL_{10} T_c} [\overline{T_c}]
 + \overline{\lambda}_{IL_{10} C} [\overline{C}] \nonumber\\ 
 &+ {{\overline{\lambda}_{IL_{10} O_x}  f_1([\overline{O_x}])}} 
- {\delta}_{IL_{10}} [\overline{IL_{10}}],
\label{eq:IL-10}
\\
\frac{d[\overline{IL_6}]}{d t}  &= 
\overline{\lambda}_{IL_6 A}  [\overline{A}]  
+ \overline{\lambda}_{IL_6 M} [\overline{M}] +{{\overline{\lambda}_{IL_6 M O_x} [\overline{M}]f_1([\overline{O_x}])  }}\nonumber\\
&+ \overline{\lambda}_{IL_6 D}  [\overline{D}] + {{\overline{\lambda}_{IL_6 O_x}  f_1([\overline{O_x}])}}
- \delta_{IL_6} [\overline{IL_6}].\label{eq:IL6}
\end{align}

\subsubsection*{Newly added compartments: ECs, VEGF and Oxygen}
ECs proliferate on their own at a very slow rate \cite{muhleder2021endothelial}. However, with VEGFs present, this growth rate increases significantly \cite{vona2009angiogenesis, cao2007angiogenesis}; so, we 
utilize a VEGF-catalyzed logistic growth for their
population. In contrast, endogenous angiogenesis inhibitors, such as IL-12, hinder EC's proliferation \cite{nyberg2005endogenous}. There are other TAFs and endogenous angiogenesis inhibitors, but we only include VEGF and IL-12 here for simplicity. We consider normoxic and hypoxic cell death for ECs. Macrophages, Adipocytes, and hypoxic Cancer cells secrete VEGF \cite{cao2007angiogenesis,fox2007breast,ferrara2002vegf}. Finally, the ECs, which form vessels, will be the oxygen source, and all the cells consume oxygen. The consumption rates are grouped by cell type to avoid complexity. 
\begin{align}
{{\frac{d[\overline{E_N}]}{dt}}}  &= {\overline{\lambda}_{VE_N}[\overline{V}]\Big(1-\frac{[\overline{E_N}]}{\overline{E_N}_0}\Big)[\overline{E_N}]}-\overline{\delta}_{E_N IL_{12}}[\overline{IL_{12}}][\overline{E_N}] \nonumber \\
&-\Big({\delta}_{E_N}+\overline{\delta}_{E_N O_x} f_2([\overline{O_x}])\Big)[\overline{E_N}], \label{ndim_eq:EN}
\\
{{\frac{d[\overline{V}]}{dt}}} &= {\overline{\lambda}_{V A}}  [\overline{A}] + {\overline{\lambda}_{V M}}  [\overline{M}] +  {{\overline{\lambda}_{VCO_x}[\overline{C}]f_1(Ox) }} \nonumber \\
&+{{\overline{\lambda}_{VO_x} f_1([\overline{O_x}])}} - {\delta_{V} [\overline{V}]}\label{ndim_eq:VEGF}
\\
{{\frac{d[\overline{Ox}]}{d\tau}}} &= \overline{\lambda}_{O_xE_N}([\overline{E_N}]) - \overline{\delta}_1 \left([\overline{T_N}]+[\overline{T_h}]+[\overline{T_c}]+ [\overline{T_r}]\right)[\overline{O_x}] \nonumber \\
&-\overline{\delta}_2 \left([\overline{D_N}]+ [\overline{D}] \right)[\overline{O_x}]-\overline{\delta}_3\left( [\overline{M_N}]+[\overline{M}]\right)[\overline{O_x}]\nonumber \\
&-\overline{\delta}_{4} [\overline{C}][\overline{O_x}]-\overline{\delta}_{5} [\overline{A}][\overline{O_x}]-\overline{\delta}_{6} [\overline{E_N}][\overline{O_x}]-\delta_{O_x}[\overline{O_x}]\label{ndim_eq:OX}
\end{align}

\subsection*{Data preparation}
We utilized RNA-sequencing data  for three mice, accessible through the Gene Expression Omnibus (GEO) database under the accession code GSE76772 \cite{cai2017transcriptomic}, encompassing four distinct tumor progression stages: hyperplasia at week 6, adenoma/MIN at week 8, early carcinoma at week 10, and late carcinoma at week 12. This data was primarily used to compare gene expression patterns across various cancer stages. Raw gene expression data was obtained through directional RNA-sequencing, and statistical techniques were employed to eliminate genes with low transcriptional activity. Then the refined data set was normalized through Differential Expression analysis for Sequence count data (DESeq) \cite{anders2010differential}.

We use a digital cytometry method of CIBERSORTx B-mode \cite{newman2019determining} to infer the different cell fractions from the RNA-sequencing data. For this task, we need a signature matrix containing information about cell-specific gene expression signatures \cite{le_review_2021,Aronow2022}. CIBERSORTx has an in-built signature matrix for homo sapiens, but we need one for mice. So, we use the ImmGen signature matrix from the Immunological Genome Project \cite{heng2008immunological}. It contains gene expression information for 207 immune cell types, most are certain immune cell subtypes consolidated as one variable in our model. For non-immune cells, a good resource is the Tabula Muris database \cite{schaum2018single}. They have a single-cell RNA-sequence (scRNA-seq) atlas for all important mouse organs and tissues. This data consists of two sets: data acquired from Fluorescence-activated cell sorting (FACS) and data acquired from Droplets. Based on the above paper, the former shows a better variety and is compatible with Immgen and CIBERSORTx. However, unlike the ImmGen data, this data cannot be readily used as a signature matrix. First, we must cross-reference the scRNA-seq count data with the FACS annotations to convert the obscure cell barcodes to meaningful cell names. Then, we upload our polished scRNA-seq data to CIBERSORTx to create a signature matrix. This will create a signature matrix to infer the non-immune cell fractions from the gene expression data.

We need a framework to estimate the number of cells from the calculated cell fractions. Sun et al. share their procedure of isolating macrophages and cancer cells for mouse breast cancer in a detailed protocol \cite{sun2022isolation}. Their samples are derived from the same mouse model used in the study from which we obtained our data. According to their protocol, they can extract $4.5\pm 0.9 \times 10^7$ cancer cells from 1 gram of tumor with more than 90\% precision. This number is obtained from two 0.5-gram tumors, each almost 10 millimeters in diameter. They mention that they avoid collecting necrotic cells. They also collected $2.2\pm 0.6 \times 10^6$ tumor-associated macrophages (TAMs) from 1 gram of the same tumor with the same precision. In a study serving as the basis for this protocol, Sun et al. also provide information on tumor volumes at various stages \cite{sun2021activating}. According to their paper, the tumor volume measures approximately 200 $mm^3$ during hyperplasia, 400 $mm^3$ during adenoma, 700 $mm^3$ during early carcinoma, and 1400 $mm^3$ during late carcinoma. These measurements are calculated in caliper format $L \times W^2 /2$, where L is the long diameter and W is the short diameter. Assuming the tumor sections are close to spheres, we can take L to be approximately equal to W. Hence, during hyperplasia, the tumor is 7.36 mm in diameter, 9.28 mm during adenoma, 11.18 mm during early carcinoma, and 14.09 mm during late carcinoma. 

% This important aspect of spatial proliferation will be considered further in future work addressing the spatiotemporal evolution of the system.

So if every 1 gram sample requires two tumors of almost 10 mm diameter to get $4.5\pm 0.9 \times 10^7$ cancer cells and $2.2\pm 0.6 \times 10^6$ TAMs, we will have the following estimates for our data:
\begin{itemize}
    \item $1.65\pm 0.33\time 10^7$ cancer cells and $8.09\pm 0.22 \times 10^5$ TAMs during hyperplasia.
    \item $2.08\pm 0.41 \times 10^7$ cancer cells and $1.02\pm 0.27 \times 10^6$ TAMs during adenoma/MIN.
    \item $2.51\pm 0.50 \times 10^7$ cancer cells and $1.22 \pm 0.33 \times 10^6$ TAMs during early carcinoma.
    \item $3.17 \pm 0.63 \times 10^7$ cancer cells and $1.54\pm 0.42 \times 10^6$ TAMs during late carcinoma.
\end{itemize}
In this study, the number of TAMs adds up to $M_N+M$ from equations \eqref{eq:MN} and \eqref{eq:M}. With the cell fraction values available 
(Figure~\ref{fig:cell-fracs}) 
and knowing the total number of macrophages, we can promptly calculate the other immune cell quantities. 

As for the non-immune cells, we use:
\begin{equation}
    \sum_{i \in I} N_i V_i + \sum_{j \in J} M_j V_j + N_C V_C = \textrm{Total Volume}. \label{eq:tot_vol}
\end{equation}
Where $N_i$ and $M_j$ are the numbers of $i$-th immune cell and $j$-th non-immune cell, respectively. The sets $I$ and $J$ contain indices for immune and non-immune cells (Adipocytes and ECs). The values $V_i$ and $V_j$ are the single cell volume of the corresponding immune and non-immune cells. The values $N_C$ and $V_C$ are the number and volume of cancer cells. Recall that the reported tumor volumes are devoid of necrotic cells. Using table \ref{tab:cellsizes} with the tumor volumes mentioned, we can get the number of non-immune cells using formula \eqref{eq:tot_vol}.

Tables \ref{tab:cellnumbers} and \ref{tab:cytonumbers} show the calculated number of cells along with the value of cytokines inferred directly from the gene expression data.  
For the oxygen levels, Ebbesen et al. report the following \cite{ebbesen2004linking}:
\begin{itemize}
    \item 10000-50000 ppm is normoxic  (equivalent to 1-5\% $O_2$ per kPa)
\item 1000-10000 ppm is mild hypoxia when HIF pathways become significant. (equivalent to 0.1-1\% $O_2$ per kPa)
\item 10-1000 ppm is severe hypoxia when cell death or cell cycle arrest is triggered (equivalent to 0.001-0.1\% $O_2$ per kPa)
\end{itemize}
Noting the oxygen molar mass is 16 g/mol, we can turn the above values into mol/L values. Cai et al., whose data we use, speculate the decrease in the expression of tricarboxylic acid cycle (TCA cycle) genes in later cancer stages is due to hypoxia. Therefore, the levels given for oxygen in table \ref{tab:cytonumbers} are picked such that the later stages go through hypoxic conditions.

All the data is non-dimensionalized before associating them with our model. The non-dimensionalization is done by dividing each cell or cytokine value in tables \ref{tab:cellnumbers} and \ref{tab:cytonumbers} by the largest value of that cell or cytokine across all samples and all stages. Also, in simulations, we assume our week 6 is the system's initial state, corresponding to $t=0$.

\begin{table}[!h]
    \centering
    
    \begin{tabular}{|c|c|c|} \hline 
         Cell type&  Length$^*$ ($\mu m)$&  Reference\\ \hline 
         Dendritic cells & 10-15 &  Tasnim et al.\cite{tasnim2018quantitative}\\ \hline 
         Macrophages & 21 & Chitu et al. \cite{chitu2011measurement} \\ \hline 
         Granulocytes and Monocytes& 13-15 & Kornmann et al. \cite{kornmann2015echogenic} \\ \hline 
         Lymphocytes (i.e., T-cells, B-cells, and NKs) & 10-15  & O'Connell et al. \cite{o2015practical}  \\ \hline 
         Stem cells& 7-17 & Pillarisetti et al. \cite{pillarisetti2009mechanical}  \\ \hline
         ECs& 20 &  Jiang et al. \cite{jiang2023matrix} \\ \hline 
         Luminal epithelial cells of mammary gland& 10-50 & Chen et al. \cite{chen2021baicalin}  \\ \hline
         Cancer cells & 12 & Laget et al. \cite{laget2017technical}\\ \hline
         Adipose-tissue derived stem cells  & 28 & Hoogduijn et al. \cite{hoogduijn2013morphology} \\ \hline
         Fibroblasts$^{**}$ & 10-20 & Turgay et al. \cite{turgay2017molecular}\\ \hline
         Adipocytes & 20 & Hagberg et al. \cite{hagberg2018flow} \\ \hline
    \end{tabular}
    \caption{Size of different cell types in mice. We use the midpoint in our calculations for the lengths with a given range. \\ \textit{* We assume the major (L) and minor (W) diameter of the cells are roughly equal and use the caliper format $L\times W^2/2(= L^3/2)$ to get the volume.} \\
    \textit{** We have not imputed cell fractions for fibroblasts. However, since most mammary stromal cells are adipocytes and fibroblasts, we use this fact to take the intersection of their approximate size (20 $\mu m$) as the size of stromal cells.}}
    \label{tab:cellsizes}
\end{table}

\begin{table}[!h]
\centering
\footnotesize
\resizebox{0.995\linewidth}{!}{
\begin{tabular}{|l|c|c|c|c|c|c|c|c|c|c|c|c|}
\hline
Sample: Stage & $T_N$ & $T_h$ & $T_C$ & $T_r$ & $D_N$ & $D$ & $M_N$ & $M$ & $C$ & $N^*$ & $A$ & $E_N$ \\
\hline
M1: W6 & 1382527 & 663086 & 228615& 92289 & 1227231 & 886534 & 389295 & 397821 & 13200000 & 6600000 & 416120 & 915268 \\
\hline
M1: W8 & 1615524 & 641814 & 265967 & 176412 & 1574935 & 1057404 & 577007 & 710944 & 16700000 & 8350000 & 1582762 & 916916 \\
\hline
M1: W10 & 2569211 & 689479 & 310204 & 292578 & 1636578 & 1328095 & 612205 & 942441 & 20100000 & 10050000 & 2763480 & 769340 \\
\hline
M1: W12 & 2199247 & 937204 & 222401 & 147089 & 2324486 & 1686078 & 787834 & 1171025 & 25400000 & 12700000 & 5222452 & 3028739 \\
\hline
M2: W6 & 1441964 & 529513 & 128158 & 79395 & 1115894 & 934727 & 450621 & 358170 & 16500000 & 8250000 & 461103 & 493492 \\
\hline
M2: W8 & 1545885 & 583058 & 589238 & 407590 & 1152659 & 1411717 & 292503 & 455659 & 20800000 & 10400000 & 1080624 & 185467 \\
\hline
M2: W10 & 1311444 & 478978 & 183104 & 69789 & 1243110 & 859137 & 430970 & 458725 & 30100000 & 15050000 & 2464786 & 931836 \\
\hline
M2: W12 & 2297591 & 426244 & 127776 & 181325 & 1396820 & 1057303 & 559029 & 562339 & 31700000 & 15850000 & 5578533 & 1469427 \\
\hline
M3: W6 & 1399198 & 430269 & 405214 & 79316 & 1021227 & 924575 & 395039 & 435495 & 19800000 & 9900000 & 333365 & 366687 \\
\hline
M3: W8 & 1467427 & 507931 & 395476 & 81323 & 1268459 & 836283 & 439574 & 575209 & 24900000 & 12450000 & 1514664 & 785294 \\
\hline
M3: W10 & 2161705 & 340691 & 272777 & 188039 & 1232740 & 1346494 & 554893 & 664238 & 25100000 & 12550000 & 2885384 & 1196948 \\
\hline
M3: W12 & 3171264 & 963737 & 381975 & 162666 & 1909543 & 1177989 & 658718 & 884036 & 38000000 & 19000000 & 5437350 & 2937968 \\
\hline
\end{tabular}}
\caption{Number of cells. The numbers in each cell are integer cutoffs of the calculated values.\\
\textit{* We assume the number of necrotic cells is half of cancer cells. This is a crude approximation based on observations from the literature about the percentage of viable space in the TME occupied by necrotic cells \cite{yamamoto2023metastasis}.
}}
\label{tab:cellnumbers}
\end{table}

\begin{table}[h]
\centering

\begin{tabular}{|l|c|c|c|c|c|c|c|}
\hline
Sample: Stage & $H$ & $IL_{12}$ & $IL_{10}$ & $IL_6$ & $V$ & $Ox$ \\
\hline
M1: W6 & 1000 & 63 & 417 & 384 & 776 & 3.125 \\ \hline
M1: W8 & 940 & 41 & 351 & 197 & 536 & 0.625 \\ \hline
M1: W10 & 1103 & 3 & 404 & 184 & 494 & 0.0625 \\ \hline
M1: W12 & 1050 & 2 & 455 & 193 & 798 & 0.000625 \\ \hline
M2: W6 & 1182 & 0 & 450 & 196 & 775 & 3.125 \\ \hline
M2: W8 & 932 & 16 & 723 & 35 & 721 & 0.625 \\ \hline
M2: W10 & 945 & 0 & 429 & 162 & 313 & 0.0625 \\ \hline
M2: W12 & 807 & 4 & 319 & 127 & 384 & 0.000625 \\ \hline
M3: W6 & 1521 & 27 & 511 & 336 & 1081 & 3.125 \\ \hline
M3: W8 & 1549 & 22 & 566 & 312 & 1054 & 0.625 \\ \hline
M3: W10 & 957 & 3 & 349 & 229 & 565 & 0.0625 \\ \hline
M3: W12 & 779 & 10 & 278 & 163 & 909 & 0.000625 \\ \hline

\end{tabular}
\caption{Cytokine levels. All the levels are, in fact, gene expression levels except for oxygen. The unit for oxygen concentration is mol/L.}
\label{tab:cytonumbers}
\end{table}

\subsection*{Parameter estimation}
We use an HGA scheme based on natural evolution for parameter estimation.  
We utilize MATLAB's HGA, which is included in the Global Optimization Toolbox. First, we initialize a loss function that calculates the distance between the predicted values and the measured data. 

We then start with a randomly produced set of parameters called the initial population. This can be constrained, biased, or a free random pick. For our problem, we do not accept negative parameter values. However, after several tests, we saw that the algorithm does not produce parameter values bigger than 2 for large upper bounds. So, we take the upper bound to be 2.

In each iteration, the algorithm leverages the individuals in the present generation to construct the succeeding population. The creation of this new population involves the following steps:
\begin{itemize}
    \item Assessing each member's fitness in the current population yields the raw fitness scores. This is done by utilizing the loss function.

    \item Scaling these raw fitness scores into a more practical range of values, referred to as expectation values.

    \item Selecting parents and individuals chosen based on their expectation values. MATLAB's default selection method constructs a line where each parent is associated with a line segment. The scaled value of the respective parent determines the length of each segment. The algorithm advances along the line in uniform steps, and at each step, it assigns a parent from the segment it arrives at. This way, we can pick out the elite children.

    \item Identifying certain individuals in the current population with lower fitness as elite members carried over to the next population.

    \item Generating offspring from the selected parents through random alterations to a single parent (mutation) or combining the attributes of a pair of parents (crossover).

    \item Substituting the existing population with the newly created offspring to constitute the subsequent generation.

    \item The algorithm stops when one of the stopping criteria is met.

    \item A local search algorithm starts using the best solution found by the genetic algorithm as the initial guess. We use the widely known MATLAB function fmincon, a gradient-based nonlinear constrained local minimizer. 
\end{itemize}
The last step is only adopted in HGA. The choice of the local minimizer depends on the problem. For the theory behind the genetic algorithm, we refer the readers to Holland's book \cite{holland1992adaptation}. Also, for a
review of the genetic algorithms and different hybrid versions, see the article by El-Mihoub et al. \cite{el2006hybrid}. 

Table 1 in the supplementary material shows the non-dimensional estimated values of the parameters for mice 1,2 and 3. As mentioned, non-dimensionalization is done by scaling each cell and cytokine level by the highest value across all three mice and all time points. This will enable us to compare the three mice and make the parameter estimation more stable. Note that we do not non-dimensionalize the time scale.

\subsection*{A discussion about the model's identifiability}

We recognize that the assumptions and connections we made with the literature to obtain the data may be optimistic.  For example, the cell sizes we used to infer the number of non-immune cells can add unforeseen uncertainties to our findings. However, our framework does result in 
plausible values of cells and cytokines, and if all of the above can be reported under the same study, 
then this can be a meaningful starting point for further studies and refinements.

Assuming all the model variables are practical outputs (i.e., are available for all times), our model and its parameters are Full Input-State-Parameter Observable (FISPO), a term coined by Villaverde et al. \cite{villaverde2019full}. One can evaluate the global structural identifiability of nonlinear systems by employing concepts from differential geometry. This assessment primarily relies on computing the rank of a specialized matrix known as the observability-identifiability matrix, formed by utilizing Lie derivatives. If the parameters are included in this computation as zero-dynamic variables, then the observability analysis of the parameters will be equivalent to their identifiability.  For more details of these methods, see \cite{villaverde2019full,villaverde2016structural,villaverde2018input,massonis2020finding,massonis2023autorepar}. 

We use the MATLAB package STRIKE-GOLDD to ascertain the global structural identifiability of our model's parameters. As previously noted, STRIKE-GOLDD declares all the model parameters structurally identifiable, assuming all our variables can be regarded as practical outcomes (i.e., data is available for them at all times).

A parameter can be globally structurally identifiable but not practically identifiable. \cite{wieland2021structural}. 
A parameter that minimizes the loss function (or maximizes the likelihood function) but has an infinite confidence interval is considered a practically non-identifiable parameter \cite{raue2009structural}. To investigate these parameters within the estimated range, we create synthetic data by perturbing the solutions in Figure \ref{fig:dynamics} by 5\%. 
Then we calculate the profile likelihood (see \cite{raue2009structural}) of each parameter, i.e., $PL(\theta_i)$. This is done by keeping the corresponding $\theta_i$ fixed and minimizing the loss function with respect to the rest of the parameters ($\theta_j$ where $j \neq i$). Next, we vary $\theta_i$ for each calculated profile likelihood and record the changes in the profile likelihood function. We then calculate a threshold for each set of estimated parameters by $L(\boldsymbol{\theta})+\Delta_\alpha$, where $L$ is the loss function used for parameter estimation and $\Delta_{\alpha}$ is the $\alpha$-quantile of the $\chi^2$ distribution with $df=1$ for pointwise confidence intervals. 
Profile likelihoods not exceeding this threshold will be labeled as practically non-identifiable within the specified parameter range. Additionally, parameters with very flat profile likelihood functions may have local structural identifiability issues around the estimated value. Refer to Figures 1, 2, and 3 in the supplementary materials for results. Within the investigated parameter neighborhoods ($[0,2*\theta_i]$), mouse 1 has 41 practically non-identifiable parameters, with 31 displaying visibly flat profiles, potentially indicating local structural non-identifiability. These numbers are lower for mice 2 and 3, which aligns with their Principle Component Analysis (PCA) count of sensitive variables.
Mouse 2 and 3 have 35 parameters that are practically non-identifiable, with 12 from mouse 2 and 13 from mouse 3 having visibly flat profile likelihoods within the studied parameter range. This shows these parameters are likely to remain non-identifiable even with increased data points or measurement quality. Notably, none of these parameters with flat likelihood profiles are identified as sensitive parameters for each mouse, suggesting their values won't significantly impact system robustness. However, we acknowledge limitations in this approach, as it relies on synthetic data and assumes all variables are always observable. However, the method can be adapted to accommodate feasible experimental situations.

\subsection*{Sensitivity analysis}

Our study employs a Principal Component Global Sensitivity Analysis (PCGSA) to scrutinize the parameter's impact on our model. To accomplish this, we utilize the Matlab package PeTTSy. The methodology for calculating sensitivity measures is derived from the approach outlined by the  PeTTSy developers in their seminal paper \cite{domijan2016pettsy}. We provide a summary of their technique here. We strongly encourage readers to consult the original study for a more comprehensive understanding and in-depth insights, as our description serves only as an overview of their detailed method.

Suppose $X_m(t,{\bf k})$ is a solution of an ODE system with $t$ being the time variable and ${\bf k} = (k_1, k_2, \cdots, k_N)$ being the parameters vector. So the i-th solution derivative with respect to the j-th parameter at a time $t$ is given by $\frac{\partial X_m(t,{\bf k})}{\partial k_j}(t)$. 
A Singular Value Decomposition (SVD) analysis is carried out to calculate the global sensitivity. This is to investigate the linearization of the mapping from perturbed parameters $\delta {\bf k} \in R^N$ to the changes they cause in the solution, namely $\delta {\bf X}$. Notice the mapping is to an infinite dimensional space of smooth functions. PeTTSy approximates these functions by evaluating them on a fine time grid (with $n$ time steps), which results in high-dimensional vectors. The final product will be a large matrix $M = \frac{\partial {\bf X}}{\partial {\bf k}}$ with $nK$ rows and $N$ columns, where $K$ is the number of our state variables. Like the state variables, we can include any other measurements, such as the sum of all the cells or the sum of all immune cells, as a single variable for the sensitivity analysis. 

The issue is that this matrix is relatively large, and investigating the parameters effect on the time-variant of a large set of solutions can be tedious. This is where the SVD and PCA are helpful. They help us focus only on a handful of Principal Components (PCs) to understand the overall parameters effect on the whole model. We start with a singular value decomposition of $M$. 
\begin{equation}
    M = U \Sigma V^T \label{eq:SVD}
\end{equation}
Where $U$ is an $nK \times nK$ orthogonal matrix, $\Sigma$ is an $nK \times N$ rectangular diagonal matrix with singular values $\sigma_i$ on its diagonal, and $V$ is a matrix whose columns form an orthonormal basis for the parameter space. The principal global sensitivity values are $S_{ij} = \sigma_i V_{ji}$. They show 
\begin{equation}
    \delta {\bf X} = \sum_{i,j} S_{ij} \delta k_j \sigma_i U_i + O(||\delta k^2 ||), \label{eq:sens1}
\end{equation}
where $\sigma_1 \geq \sigma_2 \geq \cdots \geq \sigma_N \geq 0$ are the singular values and $U_i$ (the columns of $U$) are PCs. 

For the parameter sensitivity plots, we use the values $S_{ij} = \sigma_i V_{ji}$ for each PC $U_i$. To determine how many PCs are needed to get a good sense of the system's overall sensitivity to parameters, we utilize a singular spectrum plot. We filter out the PCs corresponding to insignificant singular values. 

We also investigate the sensitivity of cancer cells and the total number of immune cells to the model parameters. Given that we only get one significant singular value for these two cases, the PC sensitivity and the classical notion of sensitivity will be synonymous according to the relationship \eqref{eq:sens1}.

\section{Results}
\subsection*{Dynamics}
\begin{figure}[H]
\centering
\includegraphics[width=\textwidth]{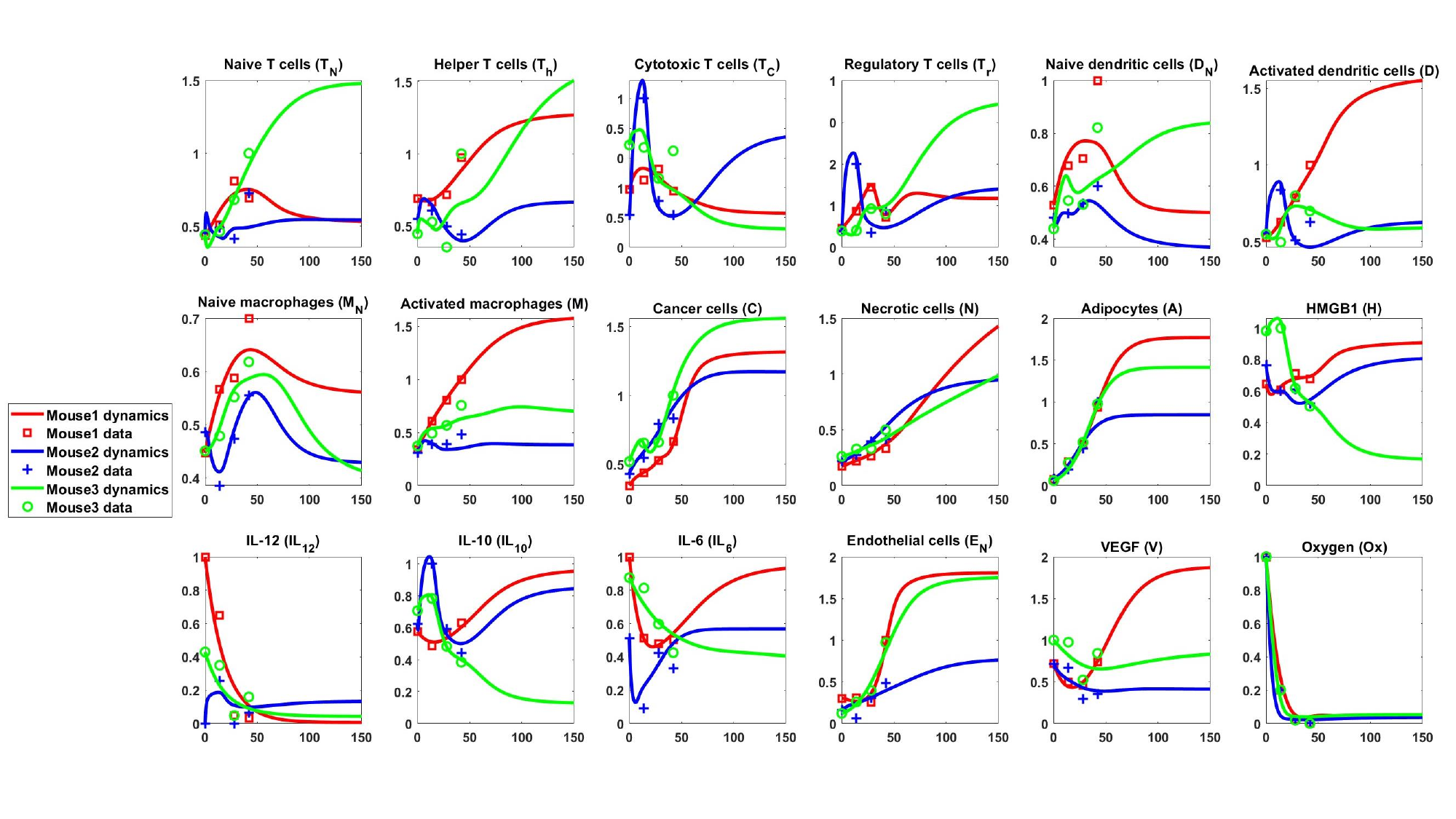}
\caption{Cells and cytokine dynamics. The x-axes show time in days. The y-axes are the dimensionless levels.} \label{fig:dynamics}
\end{figure}

We solve the non-dimensional ODEs \eqref{eq:TN}-\eqref{ndim_eq:OX} corresponding to Figure~\ref{fig:network} with parameters given in supplementary materials (Table 1). The results are shown in Figure~\ref{fig:dynamics}; the corresponding findings
illustrate a good agreement with the data. The mouse data is cut off after 12 weeks due to the subject's euthanizing, but we follow the dynamics until day 150. Moreover, the tumor samples are collected starting at week 6, which in the models marks $t=0$. Due to the system's complexity, we comment primarily on the dynamics related to the model extension, namely the ones involving ECs, VEGF, and oxygen. Further investigation will be carried out through sensitivity analysis.

Figure \ref{fig:dynamics} shows a qualitative resemblance between the macrophage and VEGF dynamics. Mouse 1 and 2 have the highest and lowest number of macrophages, respectively, consistent with adipocyte numbers, especially after $t=40$. Notice that this
is true in numerical observations, even though not manifested in the experimental data for earlier times. 
The causal effect is expected as macrophages and adipocytes are primary VEGF producers in our model, besides hypoxic cancer cells and mild hypoxia-induced pathways. Initially, VEGF levels decrease due to higher oxygen levels delaying hypoxic cancer cells and other hypoxia mechanisms contributing to its production. However, despite this decrease, endothelial cells (ECs) increase in mice 2 and 3, attributed to lower initial IL12 levels in these subjects compared to mouse 1. Consequently, even with a higher inhibition rate, ECs do not significantly decrease due to their low IL12 levels.

Having the lowest endothelial cells (ECs), mouse 2, experiences the most rapid decline in oxygen levels, maintaining the lowest oxygen values throughout the simulation. However, during the time-period from 0 to 20, where oxygen decrease is significant, the blue curve representing $E_N$ for mouse 2 surpasses that of mouse 3. This phenomenon is attributed to IL12 levels as discussed earlier. Despite starting with more oxygen producers ($E_N$), mouse 2 experiences a more pronounced oxygen decrease compared to mouse 3 due to a higher number of oxygen-consuming cell types during this short interval. In addition to ECs, mouse 2 has more T-cells, activated dendritic cells, naive macrophages, and adipocytes during this period, see Figure \ref{fig:dynamics}.

Mouse 3 has the most cancer cells, followed by Mouse 1 {for long enough times}. %Initially, Mouse 3 also retains a higher oxygen concentration. 
This could be influenced by various molecular, cellular, and hypoxic pathways, with cytotoxic T-cells enhanced by HIF pathways playing a significant role. Mouse 2, having the highest cytotoxic T-cells and the lowest adipocytes, shows the least cancer cell growth. The next section will explore more analysis of the key pathways through sensitivity analysis.

\subsection*{Sensitivity Analysis Results}
\begin{figure}[H]
    \centering
    \includegraphics[scale=0.45]{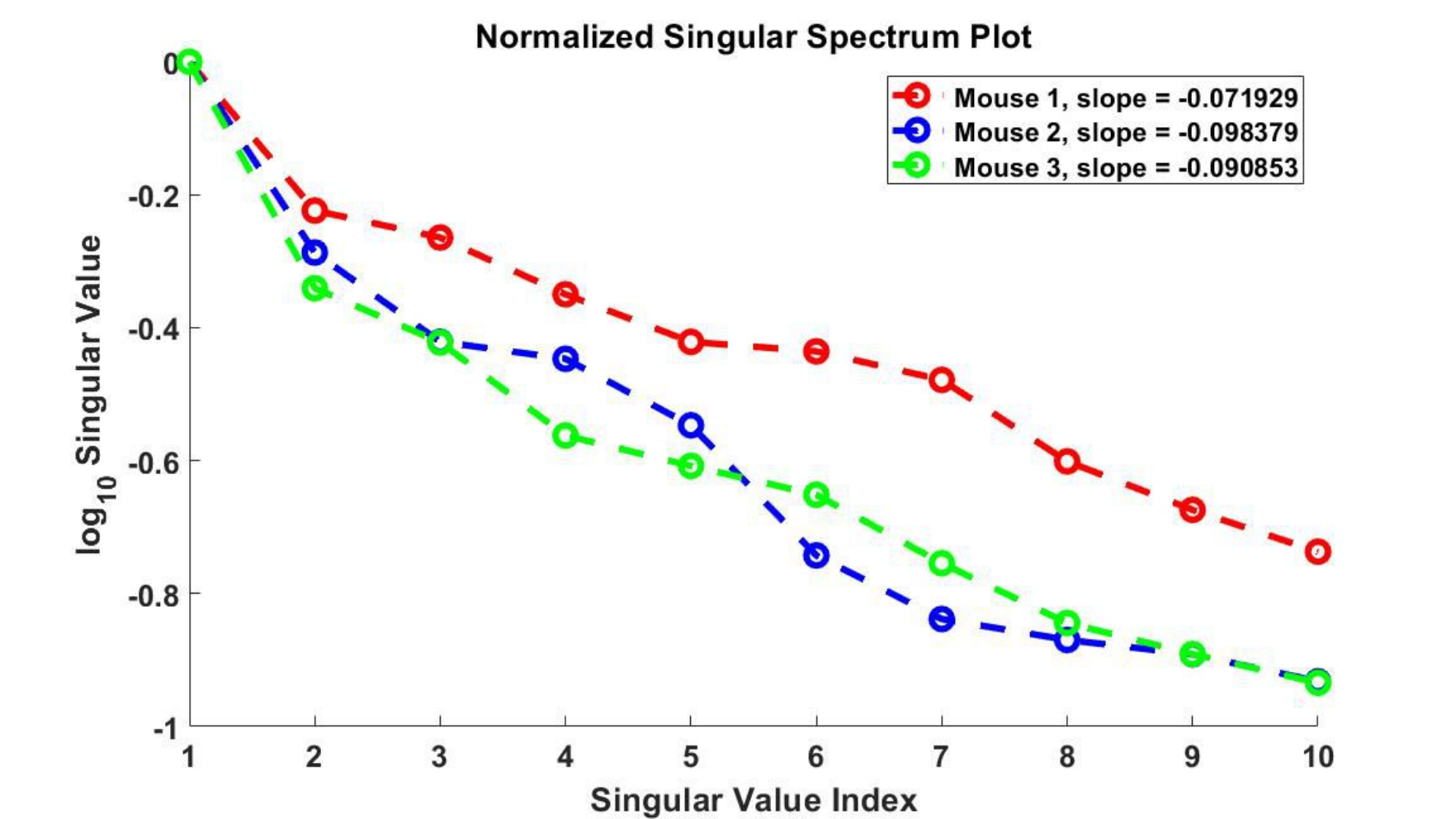}
    \caption{Singular spectrum. Top 10 largest singular values for Mouse 1,2 and 3.}
    \label{fig:SingularSpec}
\end{figure}

 We analyze the sensitivity of the entire system, cancer cells, and total immune cell count ($T_N+T_h+T_C+T_r+D_N+D+M_N+M$) to model parameters. Due to numerous state variables, we conduct a PCGSA for the first analysis. Classical global sensitivity and first PC sensitivity are equivalent for the latter two. First, we conduct a comprehensive PCGSA of the entire system relative to its parameters. With 89 parameters and 18 variables, analyzing all 18 PCs may be redundant. To determine the necessary PCs, we perform a singular spectrum analysis. Figure \ref{fig:SingularSpec} displays each mouse's top 10 normalized singular values. In mouse 1, the eighth singular value is approximately 20\% of the first. For mice 2 and 3, this occurs around the fifth and sixth singular values, respectively. Therefore, we focus on sensitivity to variations in the top 8 PCs for each mouse.

\begin{figure}[H]

    \begin{subfigure}{\textwidth}
    \hspace{0.8in}
        \includegraphics[scale=0.42]{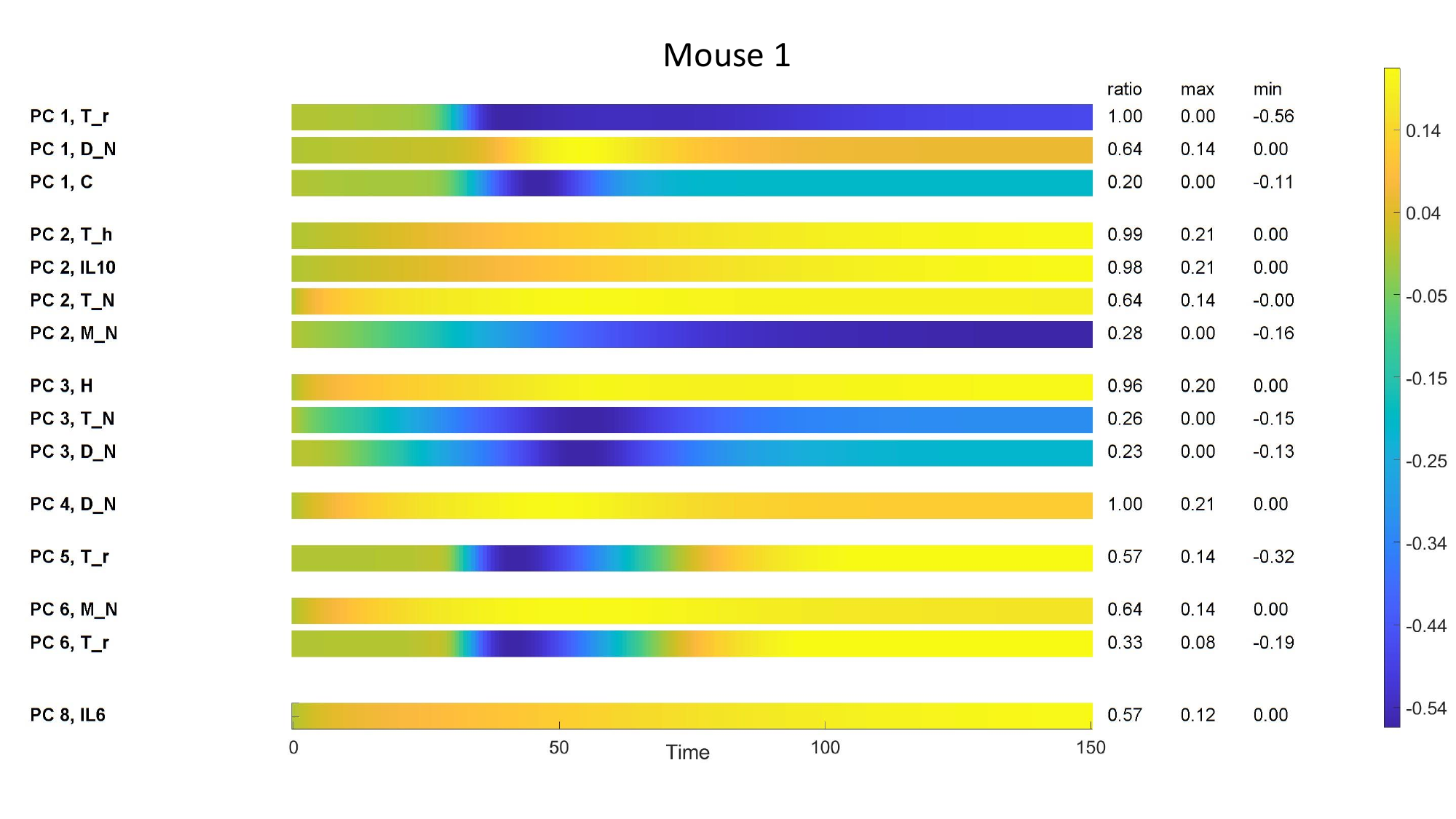}
    \end{subfigure}
    \begin{subfigure}{\textwidth}
    \hspace{0.8in}
        \includegraphics[scale=0.42]{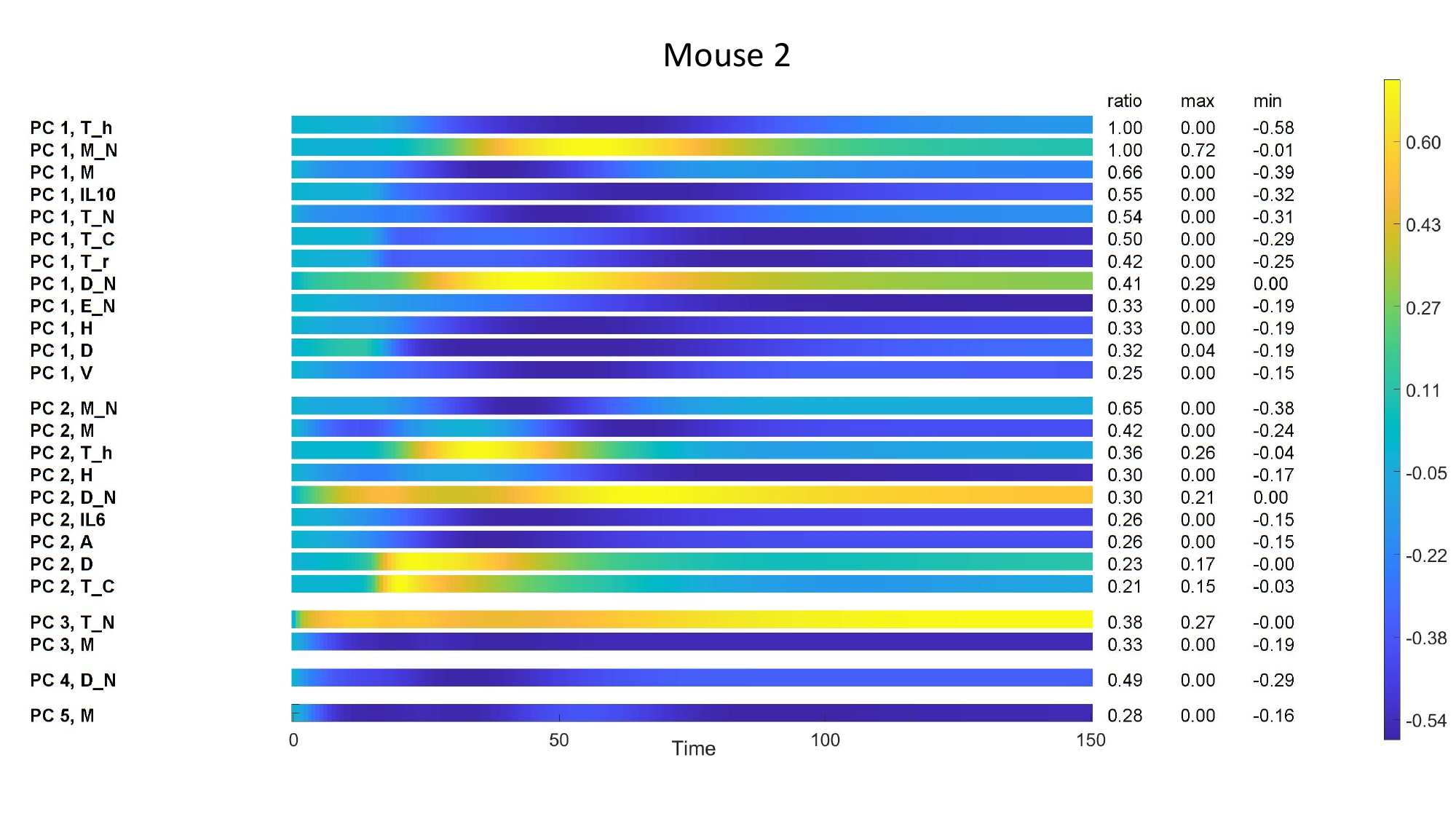}
    \end{subfigure}
    \begin{subfigure}{\textwidth}
    \hspace{0.8in}
        \includegraphics[scale=0.42]{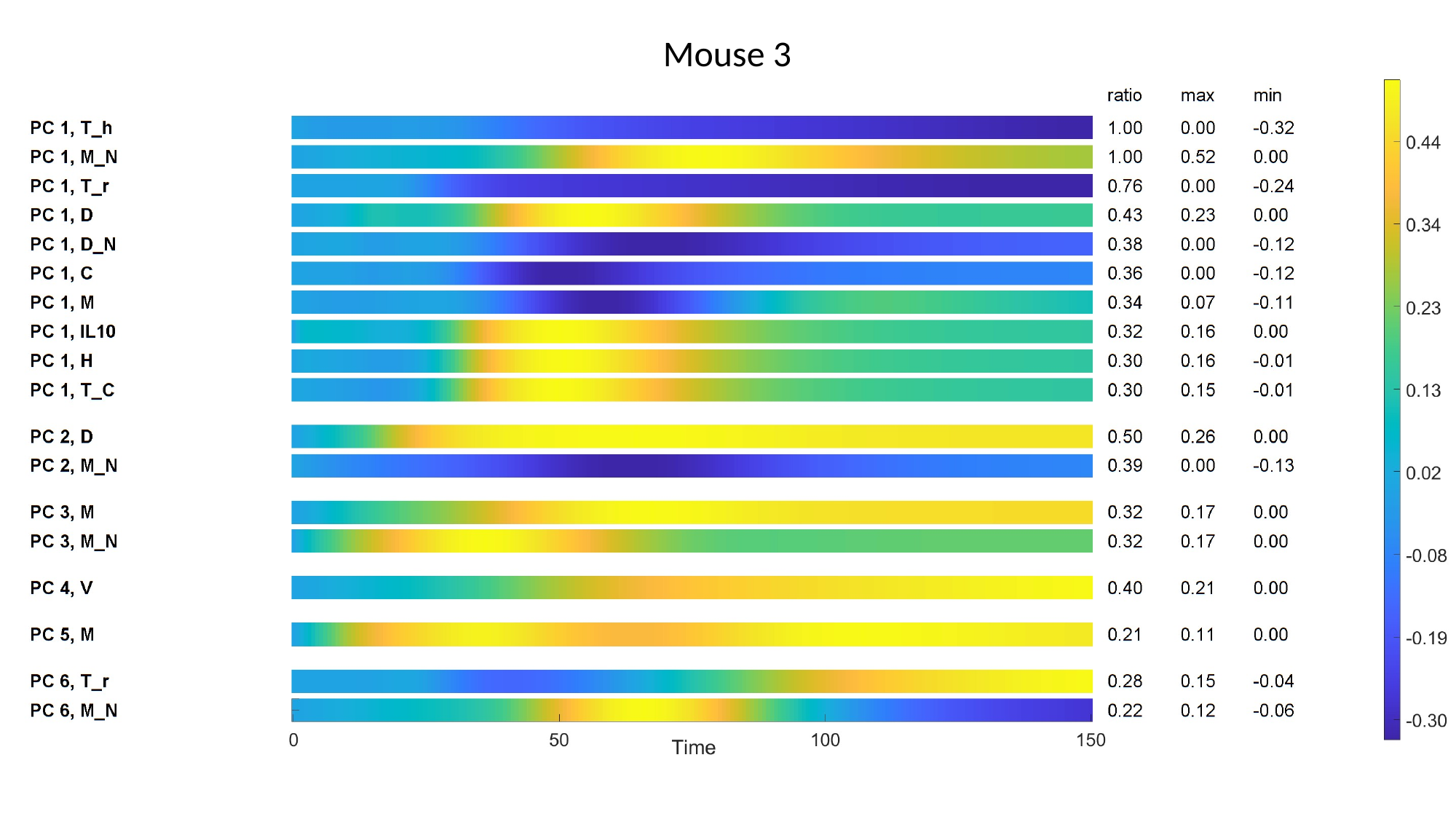}
    \end{subfigure}
    \caption{State variables sensitivity heat maps with respect to time. Only variables above 20\% of the maximum global sensitivity are shown for each mouse. For each PC, the represented variables are sorted according to their scaled ratio of change.}
    \label{fig:SSH}
\end{figure}

Figure \ref{fig:SSH} shows state variables more sensitive to parameter variations when different PCs are considered, illustrating $\sigma_i U_i$ coefficients in \eqref{eq:sens1}. Only the variables whose sensitivity exceeds 20\% of the global maximum sensitivity are included. There are sensitive state variables for all 8 PCs considered in the analysis for mouse 1. However, mice 2 and 3 only have variables exceeding 20\% of the global maximum up to the fifth and sixth PCs, respectively. This was expected based on the observations in Figure \ref{fig:SingularSpec}. We notice the dominant presence of immune cells for each PC in all three subjects. Cells such as regulatory and helper T-cells, macrophages, and dendritic cells are among the most repeated variables. These cells are all involved in producing IL10, the most sensitive cytokine in the plot. This plot can also help in designing experiments for validation. The variables shown in this plot and, more specifically, the ones shared in all three subjects are appropriate choices for experimental observations.

 \begin{figure}[H]
    \centering
    \includegraphics[scale=0.42]{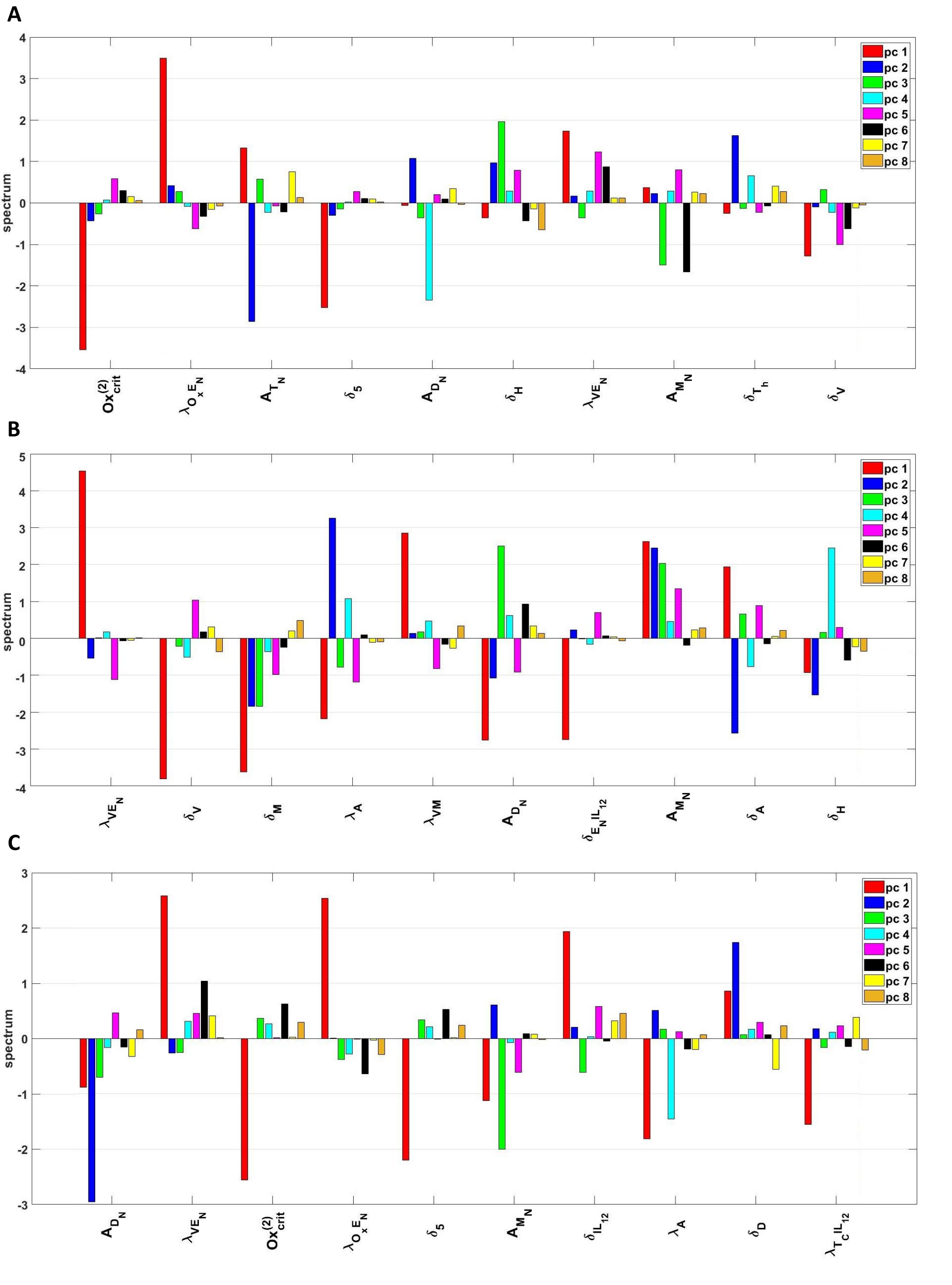}
    \caption{PCGSA of the top 10 most sensitive parameters for (A) Mouse 1, (B) Mouse 2, and (C) Mouse 3.}
    \label{fig:PPSA}
\end{figure}

Figure \ref{fig:PPSA} shows the PCGSA result for mice 1, 2, and 3. This plot has to be accompanied by Figure \ref{fig:SSH} for clearer biological inferences.
Parameters present in all three plots are $\lambda_{VE_N}$, $A_{M_N}$ and $A_{D_N}$. Changing $\lambda_{VE_N}$ inflicts the highest variation in the first PC in all three mice. Figure \ref{fig:SSH} shows immune cells such as regulatory and helper T-cells, macrophages, and dendritic cells are the most susceptible to this parameter variation. This parameter describes the promotion rate of EC proliferation by VEGF. One can connect this to the rate at which vessels are formed in the microenvironment. Also, $\lambda_{VE_N}$ involves two variables from the new model extension; see Figure~\ref{fig:network}. 

Parameters $A_{M_N}$ and $A_{D_N}$ are the innate naive macrophages and naive dendritic cell proliferation rates. 
For mouse 1, $A_{M_N}$ affects PC 6 and 3, with naive macrophages and HMGB1 receiving the highest impact. In mice 2 and 3, PCs 1, 2, 3, and 5 are affected the most, with naive and activated macrophages and helper and regulatory T-cells being the most sensitive variables to this parameter. Finally, besides the dendritic cell subtypes, helper and naive T-cells and activated and naive macrophages are included in the PCs, which are significantly influenced by $A_{D_N}$ variation. However, these effects are expected, as they control the production of two crucial naive immune cells and, consequently, their activated form. These cell types are involved in some of the most central interactions within the system. 

Oxygen-related parameters $Ox_{crit}^{(2)}$, $\lambda_{OxE_N}$, and $\delta_5$ are among the top ten most sensitive parameters in mice 1 and 3. All three show significant sensitivity for the first PC, greatly affecting helper and regulatory T-cells, macrophages, and dendritic cells. The parameter $Ox_{crit}^{(2)}$ decides the transitioning threshold from mild to severe hypoxia, $\lambda_{OxE_N}$ describes how much oxygen is provided by ECs/vessels, and $\delta_5$ is the oxygen uptake rate by adipocytes. This is interesting since in our other studies, parameters related to adipocytes had a considerable effect on the model \cite{mohammad2022investigating,mohammad2021mathematical}. 
Adipocytes are abundant in breast tissue, and adiposity is considered a well-known risk factor \cite{soguel2017adiposity}. Besides affecting cancer directly, this suggests adipocytes also affect the microenvironment immune profile through oxygen-related pathways. Moreover, $\lambda_A$ and $\delta_A$, which are adipocyte production and death rates, are placed among sensitive parameters; see Figure~\ref{fig:PPSA} and tables 2,3 and 4 in the supplementary material.

We should point out that IL-12 is affecting mouse 2's first PC by inhibiting ECs ($\delta_{E_NIL_{12}}$). It has a similar effect on mouse 3 by promoting the cytotoxic cells ($\lambda_{T_CIL_{12}}$). So, despite not being among the sensitive state variables in Figure \ref{fig:SSH}, IL-12 bears a considerable impact on the immune profile through the PC1 sensitive cells such as $T_h$, $M_N$, $M$ and $T_r$. Finally, in mouse 2, the VEGF production by macrophages ($\lambda_{VM}$) seems to have a similar effect through PC1 on the system. 

\begin{figure}[H]
    \centering
    \includegraphics[width=\textwidth]{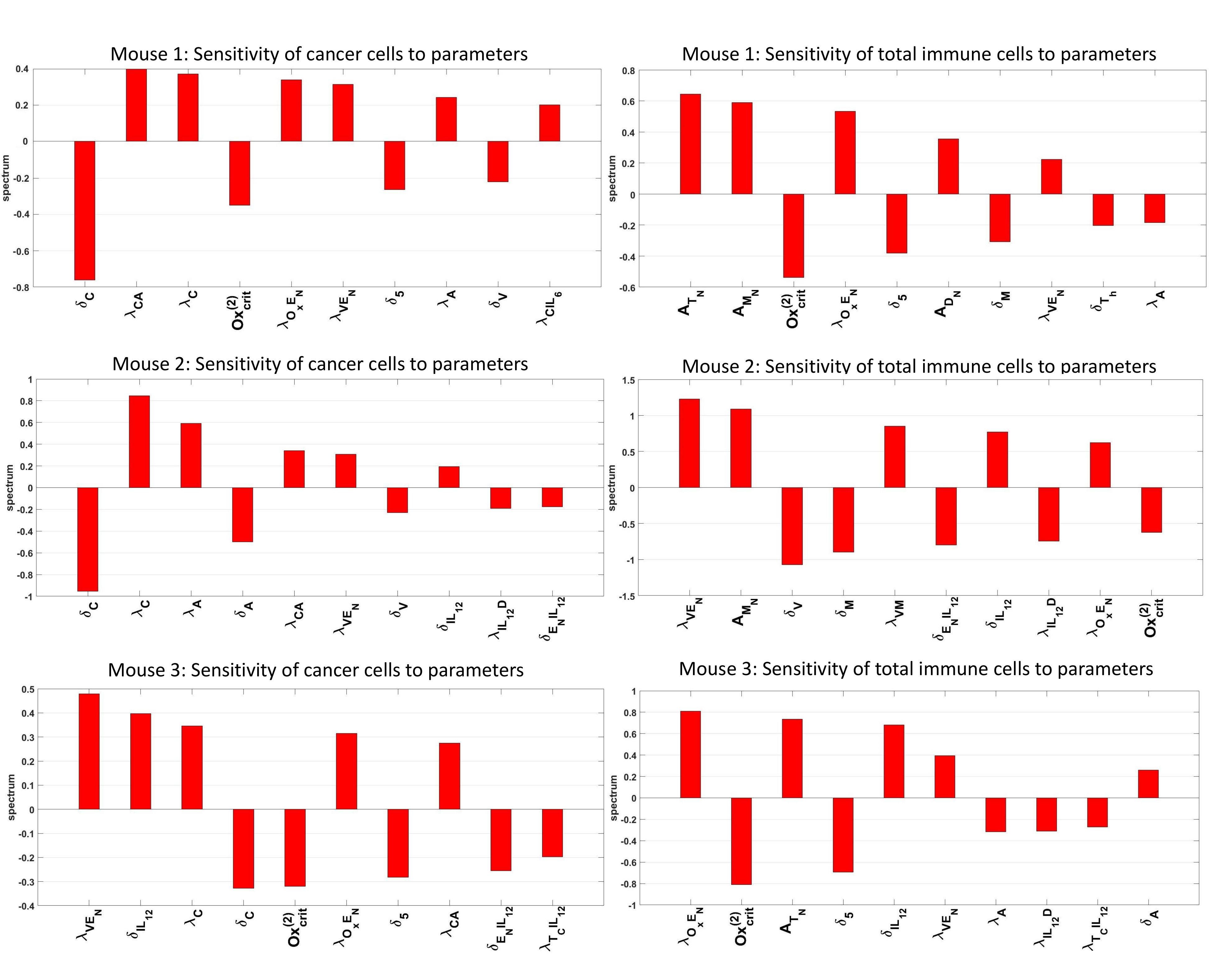}
    \caption{Sensitivity analysis with respect to cancer and total immune cells. Top 10 most sensitive parameters with respect to (left) cancer cells, (right) total immune cells.}
    \label{fig:CISA}
\end{figure}

Overall, the PCGSA shows the importance and intricacy of the immune system interactions in the TME. All parameter variations, especially parameters involving vasculature and oxygen delivery from our model extension, substantially influence crucial immune cells. We now examine whether the previously discussed parameters significantly influence diagnostic metrics, specifically the cancer and the total immune cell counts. Figure \ref{fig:CISA} illustrates the sensitivity analysis focusing on these crucial measures.

As expected, cancer levels are sensitive to parameters directly involved in 
the equation for their evolution
(see Eq.~\eqref{eq:C}) such as $\delta_C$, $\lambda_C$, $\lambda_{CA}$ and $\lambda_{CIL_6}$. This result agrees with the findings from the simpler model \cite{mohammad2022investigating}. Also, when calculating the total immune cells' sensitivity to parameters, immune cells' innate production and death rates appear among the most sensitive parameters.

In Figure \ref{fig:CISA}, $\lambda_{VE_N}$, $Ox_{crit}^{(2)}$ and $\lambda_{OxE_N}$ are the most repeated, warranting particular attention in our study. They were present in the PCA and are, importantly, a part of the model extension; see Figure~\ref{fig:network}. The results show that an increase in $\lambda_{VE_N}$ leads to an increase in all subjects' immune cells and cancer cells. This might be due to less hypoxic death for cancer cells or more pro-tumor cytokine production by immune cells such as macrophages. 
Increasing $Ox_{crit}^{(2)}$ causes a quicker transition to severe hypoxia, decreasing the total number of immune and cancer cells. This effect can be due to several factors, such as an increase in cytotoxicity or a decrease in IL-6 production. Note that the result for cancer sensitivity to $Ox_{crit}^{(2)}$ is not included in Figure~\ref{fig:CISA} for mouse 2 since it is the 13th most sensitive parameter. Nevertheless, the value is negative like the others. 

The oxygen delivery rate by the ECs/vessels ($\lambda_{OxE_N}$) is positively related to the total number of immune and cancer cells. Again, this is not included in the figure for mouse 2 since it is the 11th most sensitive parameter. 
The model extension effect on cancer cells through these parameters and the immune system showcases the importance of angiogenesis and oxygenation in the TME. This, in turn, introduces a limitation to our work regarding the cancer sensitivity to $Ox_{crit}^{(2)}$. 
Even though the hypoxic contribution to cancer proliferation through HIF pathways is included, this study considers purely oxidative phosphorylation for cells and not glycolysis as a metabolism approach. However, the metabolic switch to glycolysis under hypoxic stress is a common adaptive response from cancer cells and a challenge in cancer research \cite{deberardinis2020we}.
Therefore, increasing $Ox_{crit}^{(2)}$ might not necessarily lead to fewer cancer cells if it induces a quicker metabolic change. These intricate mechanisms require deeper analysis and a more complicated model, which we defer to future research.

The relationship between adipocyte-related parameters and cancer and total immune cells can be explained by investigating the sensitivity value signs for $\lambda_A$, $\delta_A$, $\delta_5$, and $\lambda_{CA}$. In summary, if we increase the adipocyte number through increasing $\lambda_A$ or decreasing $\delta_A$, the total number of immune cells decreases, and of cancer cells increases. Increasing adipocytes greatly increases oxygen consumption, which is lethal for all cells {affected by the hypoxic cell death}, including immune cells. However, this effect is foiled for cancer cells by the direct adipocytes effect on cancer proliferation through $\lambda_{CA}$.

An increase in VEGF leads to more immune cells and more cancer cells. This can be done through decreasing $\delta_V$ (VEGF decay rate) in mice 1 and 2 and increasing $\lambda_{VM}$ (VEGF production by macrophages) in mouse 2.
The same effects occur for mouse 3 but are not included among the top 10 in figure \ref{fig:CISA}.

We also investigate the effect of IL-12-related parameters. Inhibiting the IL-12 production (increasing $\delta_{IL12}$) leads to more immune cells and cancer cells in mice 2 and 3. Less IL-12 means less cytotoxic T-cell activation, a major cancer cell inhibitor. Also, less IL-12 means more EC proliferation (see $\delta_{E_NIL_{12}}$), which results in more oxygen, promoting the proliferation of all immune and cancer cells. See the effect of its increased production by dendritic cells ($\lambda_{IL_{12}D}$) on cancer and immune cells in mice 2 and 3 in Figure~\ref{fig:CISA}. Lastly, $\lambda_{T_CIL_{12}}$ is negatively associated with cancer cells and total immune cells in mouse 3. This is the cytotoxic T-cell activation rate by IL-12, one of the primary cancer cell inhibitors.  

\begin{figure}[H]
    \centering
    \includegraphics[width=\textwidth]{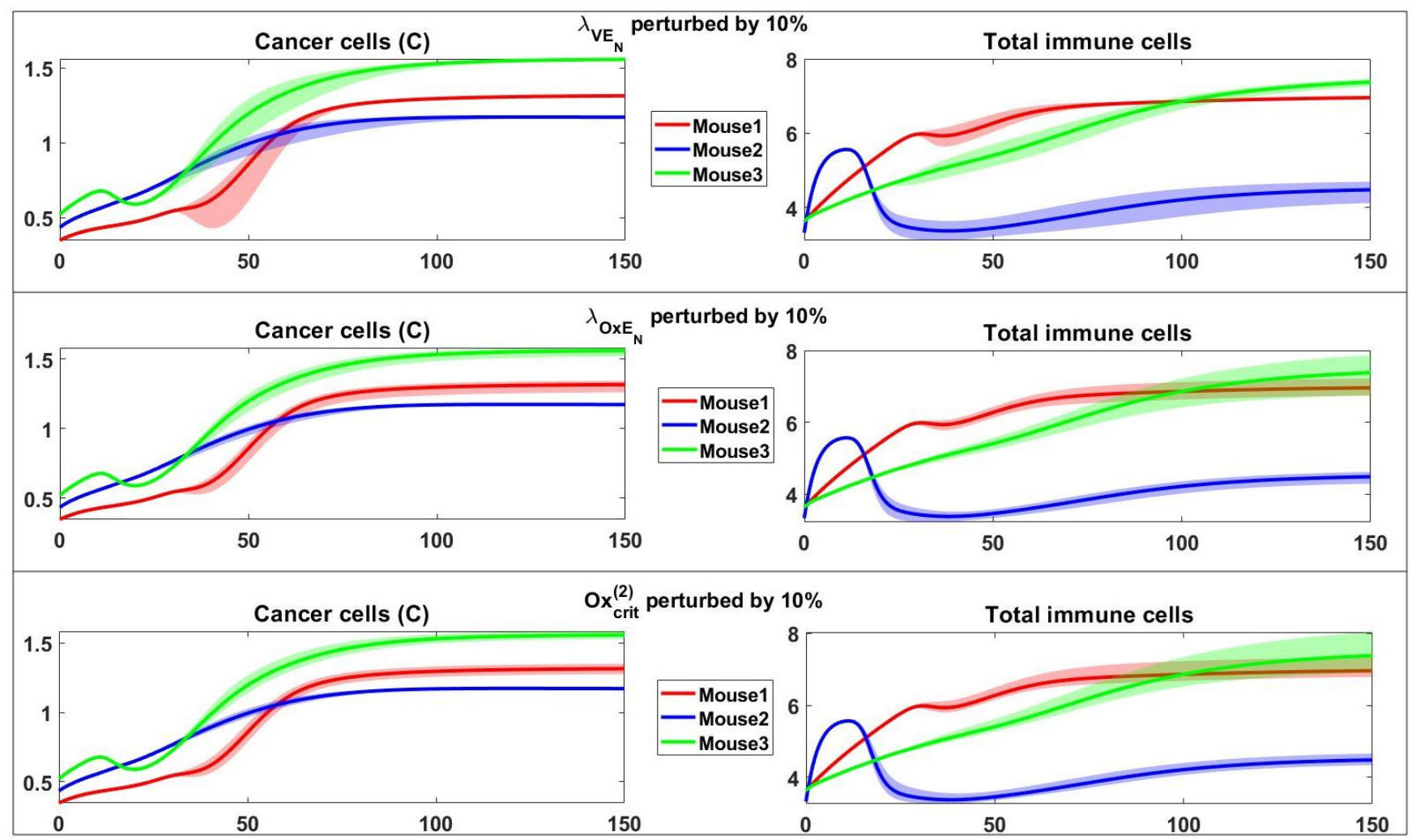}
    \caption{Effect of parameter perturbation on dynamics. The effect of perturbing $\lambda_{VE_N}$ (top), $\lambda_{OxE_N}$ (middle) and $Ox_{crit}^{(2)}$ (bottom) on dynamics of cancer cells and total immune cells.}
    \label{fig:perturbed}
\end{figure}

We finally explore the effect of perturbing $\lambda_{VE_N}$, $Ox_{crit}^{(2)}$ and $\lambda_{OxE_N}$ on the cancer dynamics and total immune cells given their presence in all three sensitivity analyses for three mice. These parameters are associated with our model extension and include the effect of all three new compartments: ECs, VEGF, and oxygen.
Figure \ref{fig:perturbed} shows the effect of the aforementioned parameters' perturbation on cancer and total immune cell dynamics. For $\lambda_{VE_N}$, the largest impact is between $t=30$ and $t=80$, which wears off as cancer approaches a steady state. Cancer steady-state translates into a state in which cancer cannot physically grow any further due to space or resources depletion.
The impact becomes visible for the total immune cells between times 20 and 30. For mouse 1, it is a short-lived effect wearing off around time 70. For mouse 3, it takes longer but eventually tapers after time 100, and Mouse 2 shows a consistent regime throughout the simulation time. This observation implies controlling cancer through controlling angiogenesis and VEGF inhibitors should happen in earlier stages to have a positive impact. Once cancer reaches its steady state, these variations do not leave a significant impact. 
It has been discussed in the literature for humans and mice that VEGF therapy for late-stage breast cancer has not been correlated with better prognoses \cite{bergers2008modes}. 

Perturbing $\lambda_{OxE_N}$ and $Ox_{crit}^{(2)}$ results in a sustainable but less significant effect. Especially in mice 1 and 3, the impact region for total immune cells increases in time. Even though controlling hypoxia or oxygen delivery is not medically feasible, numerous studies focus on targeting the HIF pathways for treatment \cite{milani2008targeting, onnis2009development, semenza2003targeting}. 

\section{Discussion}

In this study, we extended an existing ODE model, as detailed in \cite{mohammad2022investigating}, to investigate the mechanisms that promote or are angiogenesis outcomes within the PyMT-MMTV mouse breast TME. We added three important elements: ECs, VEGF, and oxygen. ECs, forming the blood vessels' infrastructure, are crucial for transporting metabolites like oxygen. VEGF, a key cytokine, initiates and accelerates angiogenesis. It is secreted by various cells, including macrophages, adipocytes, and hypoxic cancer cells. Additionally, ECs' activity is modulated by IL-12. By incorporating these components, we effectively linked them with the previously developed model \cite{mohammad2022investigating}, thereby investigating the effect of angiogenic processes and hypoxia in breast TME.

Using PyMT-MMTV mice model data and HGA parameter estimation, we approximated the ODE model parameters and simulated results for 150 days. Significant differences emerged in dynamics: mouse 1 showed the highest cancer levels, while mouse 2 showed the lowest levels attributed to cytotoxicity and adiposity. Other notable interactions included high levels of helper T-cells and dendritic cells in mouse 1, linked to HMGB1, and elevated activated macrophages due to IL-10 and helper T-cells. Also, the number of ECs correlated with VEGF levels. However, some pathways couldn't be directly inferred from the interaction network, requiring global sensitivity analysis experiments for further investigation.

Through sensitivity analyses, we observed that similar to a previous study, adipocyte-related parameters (production and death rates) were crucial to the system \cite{mohammad2022investigating}. Additionally, the oxygen consumption by adipocytes was found to be sensitive. Several studies have highlighted adipocytes' effect on hypoxia in the TME. Nieman et al. argue that in individuals with excessive adipocytes, the adipose tissue is associated with chronic hypoxia, contributing to tumor progression through HIF pathways \cite{nieman2013adipose}. Trayhurn reasons the adipocytes' sheer mass and abundance is a hypoxia-promoting factor \cite{trayhurn2013hypoxia}. 

Furthermore, a subset of parameters consistently emerged as influential factors affecting the whole system's PCs and the dynamics of cancer cells and the immune system in all subjects. These parameters, which describe the ECs promotion by VEGF ($\lambda_{VE_N}$), oxygen delivery by the ECs/vessels ($\lambda_{OxE_N}$), and the threshold value for transitioning from mild to severe hypoxia ($Ox_{crit}^{(2)}$), are all related to angiogenesis and oxygenation. The first parameter introduces an intervention opportunity for controlling cancer progression and modulating the immune response. In contrast, $\lambda_{OxE_N}$ depends more on the cells' fitness landscape and tissue content.  
In all three mice, increasing  $\lambda_{VE_N}$ led to higher cancer cell counts and greater total immune cell numbers, suggesting angiogenesis plays a crucial role in the TME. Additionally, changing the threshold $Ox_{crit}^{(2)}$ influenced the immune cell dynamics and cancer progression. These findings underscore the importance of oxygen levels in the microenvironment and suggest targeting hypoxia-related pathways could be a promising avenue for therapy. 
Recall, however, that the cancer cell 
metabolic plasticity is a relevant factor ~\cite{CELORA} that merits further study.
The oxygen delivery rate by ECs/vessels ($\lambda_{OxE_N}$) consistently impacted cancer and immune cells in all three mice. This shows the critical role of efficient oxygen transport in regulating the immune response within the TME. It also highlights the importance of angiogenesis in oxygen supply, which can affect both immune cell function and cancer cell proliferation.

We vary the three parameters discussed to study their effects on cancer and total immune cells. Perturbing $\lambda_{VE_N}$ shows a time-sensitive effect, showing the early intervention significance in managing cancer progression. The short-lived impact on total immune cells in mouse 1 and the longer-lasting effect in mouse 3 highlight the nuanced responses in different scenarios. This observation aligns with existing literature cautioning against late-stage VEGF therapy for breast cancer, emphasizing timely intervention to yield positive outcomes \cite{bergers2008modes}. Perturbing $\lambda_{OxE_N}$ and $Ox_{crit}^{(2)}$ has a less significant but sustainable effect on total immune cells, especially in mice 1 and 3 over a longer period. While directly controlling hypoxia or oxygen delivery may be challenging, the lasting impact suggests promising therapeutic avenues, as seen in studies targeting the HIF pathways for treatment. \cite{milani2008targeting, onnis2009development, semenza2003targeting}.

Our results connect with several clinical trials that have demonstrated the VEGF inhibitors efficacy, such as Bevacizumab, in breast cancer therapy \cite{miller2007paclitaxel}. These inhibitors reduce tumor vasculature, disrupt angiogenesis, and indirectly influence oxygenation levels. Importantly, this approach can potentially enhance immuno-therapies efficacy by alleviating hypoxia-induced immunosuppression \cite{jain2005normalization}. By targeting processes summarized as $\lambda_{VE_N}$ in our model, these therapeutic approaches underscore the potential
clinical relevance of our findings. They also highlight the importance of personalized treatment strategies that consider the specific angiogenesis and oxygenation dynamics in breast cancer patients. Our work paves the way for the potential in-silico exploration of
the effects of such inhibitors in future studies.

This study's findings should be interpreted considering its limitations. One major constraint is the lack of refined time-course data, which, if available, could enhance model precision and prediction reliability. Exploring spatially distributed data and comparing a spatiotemporal version of our model with such data would be extremely interesting.
Despite having a sizable system, we acknowledge some key role players in TME are not considered. This constitutes an opportunity for more extensions depending on the particular research needs. 
Additionally, numerous factors such as the role and type of macrophages~\cite{bullbyrne}, as well as the cells' phenotypic variation under different oxygen conditions~\cite{CELORA} are among factors worth further considering in the context of the present network.
Lastly, we emphasize the potential metabolic changes occurring in cancer cells, making them 
nearly immune to hypoxia. This is known as the Warburg effect \cite{liberti2016warburg} and is a HIF pathways' outcome \cite{courtnay2015cancer}. Despite including the HIF pathways' effect on the immune system, we did not consider the pathways leading to the Warburg effect for simplicity.

Future studies are needed to overcome the limitations highlighted above, like seeing how the metabolic changes in cancer can affect the system outcome. Moreover, the direct interactions between cancer cells and macrophages can be included in the model by considering the macrophages' polarization to pro and anti-tumor subtypes rather than just through the molecular pathways. Furthermore, this model lays the groundwork for developing a TME spatio-temporal PDE model. {Such a model can capture interesting phenomena such as VEGFs' chemotactic signaling to endothelial cells affecting} the vessel {network formation and} location and the oxygen penetration in the tissue influencing the dynamics and shape of the TME.

\bibliographystyle{plain}
\bibliography{main}  %%% Uncomment this line and comment out the ``thebibliography'' section below to use the external .bib file (using bibtex) .





% \end{thebibliography}

\end{document}

\maketitle
\begin{longtable}{|l|c|c|c|}
\hline
\textbf{Parameter} & \textbf{Mouse 1} & \textbf{Mouse 2} & \textbf{Mouse 3} \\
\hline
\endfirsthead
\hline
\textbf{Parameter} & \textbf{Mouse 1} & \textbf{Mouse 2} & \textbf{Mouse 3} \\
\endhead
\endfoot

$\overline{A}_{T_N}$ & 0.11494 & 0.690845 & 0.264834 \\
\hline
$\overline{\lambda}_{T_hH}$ & 0.0240562 & 0.0384676 & 0.00116581 \\
\hline
$\overline{\lambda}_{T_hD}$ & 0.0262746 & 0.136527 & 0.00146112 \\
\hline
$\overline{\lambda}_{T_hIL_{12}}$ & 0.00467816 & 0.510802 & 1.09824 \\
\hline
$\overline{\lambda}_{T_CD}$ & 0.0439357 & 0.32074 & 0.00783521 \\
\hline
$\overline{\lambda}_{T_CIL_{12}}$ & 0.158063 & 0.87611 & 0.83815 \\
\hline
$\overline{\lambda}_{T_rD}$ & 0.0548798 & 0.957692 & 0.0000996 \\
\hline
${\delta}_{T_N}$ & 0.000114152 & 0.0368966 & 0.000398804 \\
\hline
$\overline{\delta}_{T_hT_r}$ & 0.000209911 & 0.0487133 & 0.000558907 \\ \hline
$\overline{\delta}_{T_hIL_{10}}$ & 0.000376461 & 0.0843459 & 0.340433 \\ \hline
${\delta}_{T_h}$ & 0.0254337 & 0.0219081 & 0.00036025 \\ \hline

$\overline{\delta}_{T_CT_r}$ & 0.000110889 & 0.0819018 & 0.273413 \\ \hline
$\overline{\delta}_{T_CIL_{10}}$ & 0.0969647 & 0.0280879 & 0.113897 \\ \hline
${\delta}_{T_C}$ & 0.0688471 & 0.0330128 & 0.000975131 \\ \hline
${\delta}_{T_r}$ & 0.00010425 & 0.326525 & 0.0367857 \\ \hline
$\overline{A}_{D_N}$ & 0.0516926 & 0.232503 & 0.0566889 \\ \hline
$\overline{\lambda}_{DC}$ & 0.038799 & 0.0193485 & 0.000422486 \\ \hline
$\overline{\lambda}_{DH}$ & 0.0295092 & 0.714269 & 0.0539261 \\ \hline
$\overline{\delta}_{D_N}$ & 0.0256476 & 0.018035 & 0.0100014 \\ \hline
$\overline{\delta}_{DC}$ & 0.000275478 & 0.115248 & 0.0185889 \\ \hline
${\delta}_{D}$ & 0.0242571 & 0.179623 & 0.052247 \\ \hline
$\overline{A}_{M_N}$ & 0.0635285 & 0.133591 & 0.0653362 \\ \hline
$\overline{\lambda}_{MIL_{10}}$ & 0.000886774 & 0.126106 & 0.00110234 \\ \hline
$\overline{\lambda}_{MIL_{12}}$ & 0.0284695 & 0.0756649 & 0.184967 \\ \hline
$\overline{\lambda}_{MT_h}$ & 0.0502281 & 0.238148 & 0.0971966 \\ \hline
${\delta}_{M_N}$ & 0.0235334 & 0.0169597 & 0.00111295 \\ \hline
${\delta}_{M}$ & 0.029305 & 0.328197 & 0.0952367 \\ \hline
${\lambda}_C$ & 0.0853869 & 0.125587 & 0.144895 \\ \hline
$\overline{\lambda}_{CIL_6}$ & 0.0751915 & 0.0439589 & 0.0695638 \\ \hline
$\overline{\lambda}_{CA}$ & 0.107432 & 0.0807837 & 0.145561 \\ \hline
$\overline{\delta}_{CT_C}$ & 0.00428973 & 0.0111846 & 0.0788141 \\ \hline
${\delta}_C$ & 0.101099 & 0.0735727 & 0.0603353 \\ \hline
${\delta}_N$ & 0.00865431 & 0.0324343 & 0.000842819 \\ \hline
${\lambda}_A$ & 0.0883087 & 0.225724 & 0.130129 \\ \hline
${\delta}_A$ & 0.0100844 & 0.130291 & 0.0381802 \\ \hline
$\overline{\lambda}_{HD}$ & 0.0182524 & 0.0273787 & 0.000796859 \\ 
\hline
$\overline{\lambda}_{HN}$ & 0.000869656 & 0.0444178 & 0.000586598 \\ \hline
$\overline{\lambda}_{HM}$ & 0.0360787 & 0.0788468 & 0.000797632 \\ \hline
$\overline{\lambda}_{HT_C}$ & 0.151351 & 0.010657 & 1.4335 \\ \hline
$\overline{\lambda}_{HC}$ & 0.0515265 & 0.0273843 & 0.00110194 \\ \hline
${\delta}_H$ & 0.207768 & 0.160314 & 1.06705 \\ \hline
$\overline{\lambda}_{IL_{12}M}$ & 0.000101509 & 0.0172412 & 0.000684773 \\ \hline
$\overline{\lambda}_{IL_{12}D}$ & 0.000104781 & 0.290045 & 0.00293918 \\ \hline
${\delta}_{IL_{12}}$ & 0.0537936 & 1.42654 & 0.0515485 \\ \hline
$\overline{\lambda}_{IL_{10}M}$ & 0.00579803 & 0.260022 & 0.000399285 \\ \hline
$\overline{\lambda}_{IL_{10}D}$ & 0.00544879 & 0.543812 & 0.000401247 \\ \hline
$\overline{\lambda}_{IL_{10}T_r}$ & 0.000578267 & 0.591425 & 0.000662971 \\ \hline
$\overline{\lambda}_{IL_{10}T_h}$ & 0.0595401 & 0.838804 & 0.00030949 \\ \hline
$\overline{\lambda}_{IL_{10}T_C}$ & 0.00269325 & 0.0799835 & 1.19286 \\ \hline
$\overline{\lambda}_{IL_{10}C}$ & 0.000267713 & 0.0879543 & 0.000504439 \\ \hline
${\delta}_{IL_{10}}$ & 0.0984663 & 1.91009 & 1.16829 \\ \hline
$\overline{\lambda}_{IL_{6}A}$ & 0.00184929 & 0.2803 & 0.000114302 \\ \hline
$\overline{\lambda}_{IL_{6}M}$ & 0.0383226 & 0.0261256 & 0.0110877 \\ \hline
$\overline{\lambda}_{IL_{6}D}$ & 0.00413203 & 0.00848723 & 0.000250506 \\ \hline
${\delta}_{IL_6}$ & 0.0813427 & 0.50699 & 0.0215202 \\ \hline
$\overline{\lambda}_{VE_N}$ & 0.395741 & 0.124202 & 0.177899 \\ \hline
$\overline{\delta}_{E_NIL_{12}}$ & 0.175634 & 0.217809 & 0.301836 \\ \hline
${\delta}_{E_N}$ & 0.070161 & 0.001995 & 0.0051674 \\ \hline
$\overline{\lambda}_{VA}$ & 0.0538073 & 0.00210921 & 0.00957162 \\ 
\hline
$\overline{\lambda}_{VM}$ & 0.0248302 & 0.0587683 & 4.42E-05 \\ \hline
${\delta}_V$ & 0.071575 & 0.0630608 & 0.0156316 \\ \hline
$\overline{\lambda}_{T_rOx}$ & 0.000144838 & 0.015143 & 0.0895353 \\ \hline
$\overline{\lambda}_{DOx}$ & 0.0000464 & 0.0120347 & 0.047694 \\ \hline
$\overline{\delta}_{CT_COx}$ & 0.00427163 & 0.00471067 & 0.0910385 \\ \hline
$\overline{\delta}_{COx}$ & 0.39447 & 0.0667217 & 0.120738 \\ \hline
$\overline{\lambda}_{IL_{10}Ox}$ & 0.0000647 & 0.0419839 & 0.000500834 \\ \hline
$\overline{\lambda}_{IL_{6}MOx}$ & 0.000609844 & 0.0432017 & 0.000254094 \\ \hline
$\overline{\lambda}_{IL_{6}Ox}$ & 0.00537232 & 0.0187542 & 0.0000184 \\ \hline
$\overline{\lambda}_{VCOx}$ & 0.0000794 & 0.000915495 & 0.000176973 \\ \hline
$\overline{\lambda}_{VOx}$ & 0.0000551 & 0.000499566 & 0.0000365 \\ \hline
$\overline{\delta}_{T_NOx}$ & 0.000620033 & 1.59237 & 0.000554921 \\ \hline
$\overline{\delta}_{T_hOx}$ & 0.000114918 & 0.27741 & 0.000910755 \\ \hline
$\overline{\delta}_{T_COx}$ & 0.000432636 & 1.44542 & 0.374885 \\ \hline
$\overline{\delta}_{T_rOx}$ & 1.99498 & 1.86892 & 1.98791 \\ \hline
$\overline{\delta}_{D_NOx}$ & 0.000203923 & 0.022391 & 0.000627974 \\ \hline
$\overline{\delta}_{DOx}$ & 0.000423553 & 0.576513 & 0.000571083 \\ \hline
$\overline{\delta}_{E_NOx}$ & 0.000994608 & 0.0023774 & 0.00123224 \\ \hline
$\overline{\alpha}_{NC}$ & 0.124749 & 0.294812 & 0.0419279 \\ \hline
$\overline{\lambda}_{OxE_N}$ & 0.026568 & 0.0334369 & 0.0557984 \\ \hline
$\overline{\delta}_{1}$ & 0.000244366 & 0.035632 & 0.00349793 \\ \hline
$\overline{\delta}_{2}$ & 0.000229209 & 0.0334082 & 0.000918226 \\ \hline
$\overline{\delta}_{3}$ & 0.00031807 & 0.0130761 & 0.00644455 \\ \hline
$\overline{\delta}_{4}$ & 0.000137566 & 0.00813838 & 0.00446478 \\ \hline
$\overline{\delta}_{5}$ & 0.412508 & 0.576794 & 1.10035 \\ \hline
$\overline{\delta}_{6}$ & 0.189522 & 0.0584081 & 0.141885 \\ \hline
${\delta}_{ox}$ & 0.000404574 & 0.00903039 & 0.000306429 \\ \hline
\caption{Parameter values. The bar on top of the parameters means that they are non-dimensional. The parameters without the bar have the unit $\frac{1}{time}$.} \label{tab:ParVals} 
\end{longtable}

\begin{longtable}{|l|l|l|l|l|l|l|l|l|}
\hline
Parameters             & pc 1      & pc 2      & pc 3      & pc 4      & pc 5      & pc 6      & pc 7      & pc 8      \\ \hline
\endfirsthead
\hline
Parameters             & pc 1      & pc 2      & pc 3      & pc 4      & pc 5      & pc 6      & pc 7      & pc 8      \\ 
\endhead 
\endfoot
$\overline{Ox}_{crit}^{(2)}$      & -3.53614  & -0.42464  & -0.25717  & 0.069339  & 0.586031  & 0.2962    & 0.153752  & 0.061642  \\ \hline
$\overline{\lambda}_{OxE_N}$      & 3.491429  & 0.42133   & 0.270756  & -0.07906  & -0.62571  & -0.32538  & -0.15847  & -0.066    \\ \hline
$\overline{A}_{T_N}$              & 1.330189  & -2.85614  & 0.579152  & -0.22388  & -0.06915  & -0.21938  & 0.755073  & 0.125998  \\ \hline
$\overline{\delta}_{5}$           & -2.52152  & -0.29918  & -0.13836  & 0.018403  & 0.279479  & 0.111149  & 0.090404  & 0.020836  \\ \hline
$\overline{A}_{D_N}$              & -0.05484  & 1.080348  & -0.35724  & -2.34555  & 0.207771  & 0.092697  & 0.342784  & -0.03447  \\ \hline
${\delta}_H$             & -0.35624  & 0.966597  & 1.964781  & 0.292677  & 0.787067  & -0.43256  & -0.14458  & -0.64306  \\ \hline
$\overline{\lambda}_{VE_N}$       & 1.735592  & 0.162338  & -0.35982  & 0.282845  & 1.227462  & 0.872354  & 0.124946  & 0.121283  \\ \hline
$\overline{A}_M$                  & 0.373274  & 0.224109  & -1.4948   & 0.28401   & 0.806056  & -1.66055  & 0.268947  & 0.224423  \\ \hline
${\delta}_{T_h}$         & -0.25562  & 1.620359  & -0.13564  & 0.662397  & -0.22329  & -0.07382  & 0.402264  & 0.274689  \\ \hline
${\delta}_V$             & -1.27692  & -0.09356  & 0.320133  & -0.22238  & -1.0039   & -0.62617  & -0.11888  & -0.04717  \\ \hline
$\overline{\delta}_{6}$           & -1.17457  & -0.14022  & -0.07616  & 0.016211  & 0.159663  & 0.072958  & 0.044253  & 0.011349  \\ \hline
$\overline{\lambda}_{T_rD}$       & 0.47538   & 1.109756  & -0.05171  & 0.004948  & -0.05386  & 0.028627  & -0.22423  & -0.04692  \\ \hline
${\delta}_{IL_{10}}$     & -0.06214  & 0.657268  & -0.28598  & 0.468791  & -0.53874  & 0.456772  & 1.077436  & -0.43301  \\ \hline
$\overline{\lambda}_A$            & -1.01726  & -0.24869  & -0.47173  & 0.160391  & -0.337    & -0.03748  & -0.34017  & -0.24213  \\ \hline
${\delta}_C$             & -0.31393  & 0.21572   & 0.986364  & -0.43143  & 0.259619  & -0.37258  & 0.560911  & 0.127443  \\ \hline
${\delta}_{IL_6}$        & -0.11283  & 0.209907  & 0.577614  & -0.05874  & -0.19696  & 0.410474  & -0.03989  & 0.923171  \\ \hline
$\overline{\lambda}_{IL_{10}T_h}$ & 0.051948  & -0.54705  & 0.238322  & -0.39088  & 0.448986  & -0.37966  & -0.89852  & 0.360929  \\ \hline
$\overline{\delta}_{T_rOx}$       & -0.88415  & -0.13559  & -0.17127  & 0.050965  & 0.123088  & 0.108321  & -0.01031  & 0.005096  \\ \hline
${\delta}_{E_N}$         & -0.87129  & -0.0857   & 0.134567  & -0.11199  & -0.47534  & -0.34201  & -0.04463  & -0.04645  \\ \hline
${\delta}_{M}$           & -0.33896  & 0.424506  & 0.848067  & 0.090952  & -0.27389  & 0.409225  & -0.16943  & 0.466062  \\ \hline
${\delta}_{D_N}$         & 0.017882  & -0.35103  & 0.115024  & 0.739738  & -0.06414  & -0.02814  & -0.10645  & 0.010304  \\ \hline
$\overline{\lambda}_{IL_{6}M}$    & 0.082515  & -0.16581  & -0.43534  & 0.023124  & 0.166517  & -0.35323  & 0.059674  & -0.73357  \\ \hline
${\delta}_{D}$           & -0.15198  & -0.53495  & 0.721274  & 0.714989  & 0.094268  & -0.23342  & 0.374996  & 0.164614  \\ \hline
$\overline{\lambda}_{T_hD}$       & -0.04007  & -0.72022  & -0.01046  & -0.45623  & 0.17914   & 0.10633   & -0.40639  & -0.22383  \\ \hline
$\overline{\lambda}_{MT_h}$       & 0.272518  & -0.6814   & -0.10995  & -0.26262  & -0.11688  & 0.65392   & 0.015033  & -0.71598  \\ \hline
$\overline{\lambda}_{VA}$         & 0.642756  & 0.042043  & -0.14517  & 0.094618  & 0.442723  & 0.246291  & 0.056421  & 0.000546  \\ \hline
$\overline{\lambda}_{HC}$         & 0.108797  & -0.30031  & -0.62269  & -0.11375  & -0.24821  & 0.110875  & 0.074502  & 0.217249  \\ \hline
${\delta}_{IL_{12}}$     & 0.611503  & -0.16284  & 0.120957  & 0.052808  & 0.595032  & 0.173647  & 0.05003   & 0.085451  \\ \hline
$\overline{\delta}_{E_NIL_{12}}$  & -0.5859   & -0.0521   & 0.159247  & -0.12094  & -0.52528  & -0.37309  & -0.05544  & -0.05396  \\ \hline
$\overline{\lambda}_{T_CD}$       & -0.30037  & 0.578011  & -0.54152  & 0.01374   & -0.18148  & 0.207569  & 0.10437   & 0.067189  \\ \hline
$\overline{\lambda}_{HT_C}$       & 0.111519  & -0.2956   & -0.57185  & -0.04491  & -0.2321   & 0.17714   & -0.01492  & 0.161215  \\ \hline
$\overline{\lambda}_{VM}$         & 0.527954  & 0.040417  & -0.13205  & 0.093471  & 0.420967  & 0.270549  & 0.048285  & 0.024107  \\ \hline
$\overline{\lambda}_{HM}$         & 0.093229  & -0.25376  & -0.52502  & -0.08978  & -0.20814  & 0.100915  & 0.055966  & 0.177474  \\ \hline
$\overline{\lambda}_C$            & 0.156763  & -0.10421  & -0.51237  & 0.259611  & -0.10387  & 0.224173  & -0.32057  & -0.10375  \\ \hline
$\overline{\lambda}_{CA}$         & 0.154142  & -0.12128  & -0.49538  & 0.165588  & -0.12561  & 0.149596  & -0.24054  & -0.07135  \\ \hline
${\delta}_{M_N}$         & -0.09905  & -0.06413  & 0.432603  & -0.07988  & -0.22755  & 0.494023  & -0.07959  & -0.06442  \\ \hline
$\overline{\lambda}_{DC}$         & 0.150278  & 0.011746  & -0.4474   & 0.410812  & -0.14993  & 0.16414   & -0.45574  & -0.12613  \\ \hline
$\overline{\lambda}_{T_hH}$       & -0.01692  & -0.4439   & -0.00298  & -0.28467  & 0.107835  & 0.073648  & -0.26685  & -0.14652  \\ \hline
$\overline{\lambda}_{DH}$         & 0.097981  & 0.01323   & -0.30332  & 0.297459  & -0.09355  & 0.12661   & -0.31714  & -0.08501  \\ \hline
${\delta}_A$             & 0.299076  & 0.067247  & 0.130105  & -0.04307  & -0.04018  & -0.05007  & 0.037775  & 0.031377  \\ \hline
$\overline{\lambda}_{CIL_6}$      & 0.083261  & -0.05952  & -0.27889  & 0.133152  & -0.06553  & 0.112065  & -0.16784  & -0.04954  \\ \hline
${\delta}_{T_C}$         & -0.05609  & 0.176483  & 0.266119  & 0.007404  & 0.17341   & -0.10178  & -0.21554  & -0.06585  \\ \hline
$\overline{\lambda}_{HD}$         & 0.046444  & -0.12677  & -0.2621   & -0.04475  & -0.10438  & 0.050407  & 0.027603  & 0.089251  \\ \hline
$\overline{\delta}_{T_CIL_{10}}$  & -0.051    & 0.158987  & 0.257645  & 0.015206  & 0.164794  & -0.08812  & -0.21317  & -0.07141  \\ \hline
$\overline{\lambda}_{T_CIL_{12}}$ & -0.10375  & 0.197511  & -0.15477  & 0.069252  & -0.14733  & 0.00588   & 0.010033  & 0.028485  \\ \hline
$\overline{\delta}_{COx}$         & -0.03656  & 0.024855  & 0.105092  & -0.03445  & 0.026246  & -0.0334   & 0.049959  & 0.010544  \\ \hline
$\overline{\lambda}_{IL_{10}M}$   & 0.006444  & -0.06522  & 0.02605   & -0.04492  & 0.05056   & -0.04308  & -0.10178  & 0.041508  \\ \hline
$\overline{\lambda}_{IL_{10}D}$   & 0.005927  & -0.06028  & 0.024383  & -0.04171  & 0.047132  & -0.04012  & -0.09469  & 0.038521  \\ \hline
$\overline{\lambda}_{MIL_{12}}$   & 0.064627  & -0.02547  & -0.0232   & 0.002346  & 0.027545  & 0.089847  & -0.01142  & -0.0389   \\ \hline
$\overline{\lambda}_{IL_{6}D}$    & 0.008869  & -0.0177   & -0.04671  & 0.002751  & 0.017631  & -0.03739  & 0.006045  & -0.07824  \\ \hline
$\overline{\lambda}_{IL_{6}Ox}$   & 0.008992  & -0.01715  & -0.04549  & 0.002605  & 0.017104  & -0.03508  & 0.005427  & -0.07552  \\ \hline
$\overline{\alpha}_{NC}$          & -0.00763  & -0.01397  & -0.03856  & 0.009607  & 0.014524  & 0.004915  & -0.01612  & -0.04996  \\ \hline
$\overline{\lambda}_{IL_{6}A}$    & 0.00378   & -0.00868  & -0.02143  & -0.00012  & 0.009245  & -0.02048  & 0.004861  & -0.03858  \\ \hline
${\delta}_N$             & 0.003286  & 0.005369  & 0.014525  & -0.0035   & -0.00594  & -0.00161  & 0.005517  & 0.019057  \\ \hline
$\overline{\delta}_{T_hIL_{10}}$  & -0.00269  & 0.018346  & -0.00156  & 0.007312  & -0.00245  & -0.00034  & 0.004212  & 0.002827  \\ \hline
$\overline{\delta}_{CT_C}$        & -0.00511  & 0.003253  & 0.016117  & -0.00815  & 0.003634  & -0.00704  & 0.010054  & 0.002712  \\ \hline
$\overline{\lambda}_{T_hIL_{12}}$ & 0.002698  & -0.01215  & 0.001379  & -0.00889  & 0.003921  & 0.005174  & -0.01291  & -0.00661  \\ \hline
$\overline{\lambda}_{IL_{10}T_C}$ & 0.000312  & -0.00531  & 0.003978  & -0.005    & 0.006396  & -0.00519  & -0.01256  & 0.004642  \\ \hline
$\overline{\delta}_{CT_COx}$      & -0.00395  & 0.002512  & 0.01168   & -0.00493  & 0.002498  & -0.0046   & 0.006478  & 0.001806  \\ \hline
$\overline{\lambda}_{IL_{6}MOx}$  & 0.001177  & -0.00253  & -0.00641  & 8.74E-05  & 0.002672  & -0.00576  & 0.001246  & -0.01127  \\ \hline
$\overline{\lambda}_{MIL_{10}}$   & 0.00353   & -0.00866  & -0.00142  & -0.00331  & -0.00141  & 0.008333  & 0.000204  & -0.00908  \\ \hline
$\overline{\delta}_{DC}$          & -0.00173  & -0.00564  & 0.007982  & 0.008992  & 0.001344  & -0.00214  & 0.003845  & 0.001787  \\ \hline
$\overline{\delta}_{T_hT_r}$      & -0.00115  & 0.007688  & -0.00065  & 0.003115  & -0.00101  & -0.00023  & 0.001829  & 0.001253  \\ \hline
$\overline{\lambda}_{HN}$         & 0.001258  & -0.00348  & -0.00757  & -0.00154  & -0.00309  & 0.001074  & 0.001219  & 0.002871  \\ \hline
$\overline{\lambda}_{IL_{10}T_r}$ & 0.000239  & -0.00269  & 0.001301  & -0.00202  & 0.002355  & -0.00198  & -0.00472  & 0.001869  \\ \hline
$\overline{\lambda}_{IL_{12}D}$   & -0.00427  & 0.002074  & -0.00315  & -0.00063  & -0.00159  & 0.00148   & 0.000573  & -0.00047  \\ \hline
$\overline{Ox}_{crit}^{(1)}$           & 0.00148   & -0.00051  & -0.003    & 0.000919  & 0.001537  & -0.00018  & -0.00059  & -0.00425  \\ \hline
$\overline{\lambda}_{IL_{12}M}$   & -0.00412  & 0.002035  & -0.00311  & -0.0006   & -0.00148  & 0.0015    & 0.000571  & -0.00045  \\ \hline
$\overline{\lambda}_{IL_{10}C}$   & 0.000264  & -0.00259  & 0.000984  & -0.00174  & 0.001952  & -0.00166  & -0.00392  & 0.001608  \\ \hline
$\overline{\delta}_{1}$           & -0.00371  & -0.0004   & 0.000238  & -0.00027  & -0.00097  & -0.0008   & -6.80E-05 & -0.00017  \\ \hline
$\overline{\delta}_{3}$           & -0.00353  & -0.00039  & 0.000122  & -0.00019  & -0.00059  & -0.00053  & -1.51E-05 & -0.00011  \\ \hline
${\delta}_{Ox}$          & -0.00243  & -0.00026  & 0.000173  & -0.00019  & -0.00069  & -0.00057  & -5.34E-05 & -0.00013  \\ \hline
$\overline{\delta}_{2}$           & -0.00235  & -0.00026  & 0.000101  & -0.00014  & -0.00046  & -0.0004   & -1.93E-05 & -8.67E-05 \\ \hline
$\overline{\lambda}_{T_rOx}$      & 0.000957  & 0.002347  & -0.00016  & 6.87E-05  & -2.98E-06 & 0.000136  & -0.00045  & -5.79E-05 \\ \hline
${\delta}_{T_N}$         & -0.00083  & 0.001792  & -0.00037  & 0.000156  & 2.65E-06  & 0.000119  & -0.00048  & -7.46E-05 \\ \hline
${\delta}_{T_r}$         & -0.00134  & -0.00019  & -0.00025  & 7.97E-05  & 0.000104  & 0.000117  & -5.51E-05 & -1.82E-05 \\ \hline
$\overline{\lambda}_{VOx}$        & 0.000871  & 6.67E-05  & -0.00021  & 0.000148  & 0.000675  & 0.000431  & 7.69E-05  & 3.50E-05  \\ \hline
$\overline{\lambda}_{IL_{10}Ox}$  & 4.98E-05  & -0.00052  & 0.000218  & -0.00037  & 0.000414  & -0.00035  & -0.00084  & 0.00034   \\ \hline
$\overline{\lambda}_{VCOx}$       & 0.000756  & 5.05E-05  & -0.00017  & 0.000114  & 0.00053   & 0.000301  & 6.63E-05  & 4.29E-06  \\ \hline
$\overline{\delta}_{4}$           & -0.00069  & -8.02E-05 & -1.60E-05 & -1.00E-05 & 7.97E-06  & -1.90E-05 & 1.45E-05  & -6.40E-06 \\ \hline
$\overline{\lambda}_{DOx}$        & 0.000188  & 2.22E-05  & -0.00057  & 0.000524  & -0.00016  & 0.00024   & -0.00056  & -0.00017  \\ \hline
$\overline{\delta}_{DOx}$         & -9.29E-05 & -0.0003   & 0.00042   & 0.000445  & 5.67E-05  & -0.00013  & 0.000201  & 9.61E-05  \\ \hline
$\overline{\delta}_{T_NOx}$       & -0.00015  & 0.000319  & -7.95E-05 & 2.11E-05  & -1.69E-06 & 3.82E-05  & -8.52E-05 & -1.65E-05 \\ \hline
$\overline{\delta}_{E_NOx}$       & -0.0003   & -3.08E-05 & 2.56E-05  & -2.43E-05 & -0.0001   & -7.33E-05 & -7.39E-06 & -7.34E-06 \\ \hline
$\overline{\delta}_{T_hOx}$       & -3.35E-05 & 0.000252  & -2.13E-05 & 0.000101  & -3.04E-05 & -1.31E-06 & 5.46E-05  & 3.86E-05  \\ \hline
$\overline{\delta}_{T_CT_r}$      & -4.80E-05 & 0.000152  & 0.000235  & 9.66E-06  & 0.000152  & -8.56E-05 & -0.00019  & -5.82E-05 \\ \hline
$\overline{\delta}_{D_NOx}$       & 7.13E-06  & -7.96E-05 & 2.14E-05  & 0.000182  & -1.25E-05 & -1.80E-06 & -3.11E-05 & -8.60E-07 \\ \hline
$\overline{\delta}_{T_COx}$       & -1.04E-05 & 2.80E-05  & 4.93E-05  & 3.49E-06  & 2.78E-05  & -1.65E-05 & -4.08E-05 & -1.16E-05 \\ \hline
\caption{Full principle component sensitivity results for Mouse 1.}
\end{longtable}
\vspace{0.6in}
\begin{longtable}{|l|l|l|l|l|l|l|l|l|}
\hline
Parameters             & pc 1      & pc 2      & pc 3      & pc 4      & pc 5      & pc 6      & pc 7      & pc 8      \\ \hline
\endfirsthead
\hline
Parameters             & pc 1      & pc 2      & pc 3      & pc 4      & pc 5      & pc 6      & pc 7      & pc 8      \\ 
\endhead 
\endfoot
\hline
$\overline{\lambda}_{VE_N}$       & 4.541017 & -0.53847 & 0.015424 & 0.175511 & -1.11219 & -0.06617 & -0.05825  & 0.02015  \\ \hline
${\delta}_V$             & -3.81021 & -0.00391 & -0.21452 & -0.51542 & 1.032161 & 0.175914 & 0.308967  & -0.35903 \\ \hline
${\delta}_{M}$           & -3.62047 & -1.83484 & -1.83563 & -0.36099 & -0.98212 & -0.24535 & 0.204925  & 0.488976 \\ \hline
$\overline{\lambda}_A$            & -2.17893 & 3.262126 & -0.77578 & 1.079412 & -1.17987 & 0.100345 & -0.10553  & -0.08705 \\ \hline
$\overline{\lambda}_{VM}$         & 2.847654 & 0.136716 & 0.182879 & 0.474816 & -0.81255 & -0.16002 & -0.26364  & 0.342504 \\ \hline
$\overline{A}_{D_N}$              & -2.75649 & -1.07619 & 2.503844 & 0.615239 & -0.91772 & 0.925714 & 0.338476  & 0.131557 \\ \hline
$\overline{\delta}_{E_NIL_{12}}$  & -2.73712 & 0.225269 & -0.01675 & -0.16449 & 0.697809 & 0.066882 & 0.039207  & -0.06788 \\ \hline
$\overline{A}_M$                  & 2.631197 & 2.45075  & 2.034162 & 0.455082 & 1.351683 & -0.1828  & 0.228192  & 0.286397 \\ \hline
${\delta}_A$             & 1.943993 & -2.56786 & 0.665939 & -0.76825 & 0.883529 & -0.1434  & 0.052977  & 0.211181 \\ \hline
${\delta}_H$             & -0.92914 & -1.53671 & 0.162531 & 2.44458  & 0.295055 & -0.5924  & -0.23223  & -0.3451  \\ \hline
${\delta}_{IL_{12}}$     & 2.312262 & -0.0364  & -0.30846 & 0.278032 & -0.47436 & -0.36773 & 0.469818  & 0.124798 \\ \hline
$\overline{\lambda}_{IL_{12}D}$   & -2.24896 & 0.047751 & 0.296652 & -0.26091 & 0.456917 & 0.352959 & -0.45831  & -0.11559 \\ \hline
$\overline{\lambda}_{OxE_N}$      & 2.159935 & -0.69246 & 0.047113 & -0.34055 & -0.24168 & 0.377771 & 0.053357  & -0.41627 \\ \hline
$\overline{A}_{T_N}$              & 2.080343 & -0.04411 & -2.1378  & 0.973935 & 1.207342 & 0.711688 & 0.345328  & -0.00087 \\ \hline
$Ox_{crit}^{(2)}$      & -2.12938 & 0.690624 & -0.05531 & 0.357842 & 0.232289 & -0.38804 & -0.05326  & 0.432216 \\ \hline
$\overline{\lambda}_{DH}$         & 0.340418 & 0.89836  & -0.41295 & -1.61687 & 0.135966 & -0.81012 & -0.59315  & -0.39086 \\ \hline
$\overline{\delta}_{5}$           & -1.56614 & 0.500299 & -0.03783 & 0.246837 & 0.179956 & -0.27552 & -0.03742  & 0.308396 \\ \hline
${\delta}_{D}$           & 1.183737 & -0.04372 & -0.94972 & 0.506701 & 0.364359 & -0.01211 & 0.157653  & 0.120318 \\ \hline
$\overline{\lambda}_{T_rD}$       & -0.87081 & 0.003422 & 0.977004 & -0.3706  & -0.6141  & -0.67368 & 0.355374  & -0.14691 \\ \hline
$\overline{\lambda}_C$            & 0.885491 & 0.721519 & -0.47743 & -0.61978 & -0.25214 & 0.280742 & 0.38965   & 0.064398 \\ \hline
${\delta}_{IL_{10}}$     & -0.30398 & 0.107328 & -0.1359  & -0.01659 & 0.199833 & 0.328362 & -0.7662   & 0.718084 \\ \hline
$\overline{\lambda}_{MT_h}$       & 0.696143 & -0.25074 & -0.03793 & -0.02692 & -0.17748 & 0.22911  & -0.24077  & -0.42958 \\ \hline
${\delta}_C$             & -0.69365 & -0.42423 & 0.481903 & 0.148401 & 0.076551 & -0.02585 & -0.23646  & 0.016324 \\ \hline
$\overline{\lambda}_{HM}$         & 0.262382 & 0.42048  & -0.04928 & -0.69171 & -0.07252 & 0.144976 & 0.055975  & 0.080775 \\ \hline
$\overline{\lambda}_{HN}$         & 0.268557 & 0.43765  & -0.03859 & -0.64107 & -0.11255 & 0.200613 & 0.096427  & 0.125382 \\ \hline
$\overline{\delta}_{DC}$          & 0.571264 & 0.062888 & -0.61403 & 0.339881 & 0.263038 & -0.01315 & 0.087162  & 0.088803 \\ \hline
$\overline{\alpha}_{NC}$          & 0.24241  & 0.534567 & -0.08232 & -0.60252 & -0.2446  & 0.409767 & 0.263774  & 0.148529 \\ \hline
$\overline{\lambda}_{HC}$         & 0.248181 & 0.390166 & -0.04001 & -0.59889 & -0.09059 & 0.159219 & 0.076827  & 0.096388 \\ \hline
$\overline{\lambda}_{T_hD}$       & 0.563684 & -0.17821 & 0.04847  & 0.003819 & -0.1052  & 0.345623 & -0.47824  & -0.0345  \\ \hline
${\delta}_{IL_6}$        & -0.06212 & -0.42863 & 0.162813 & -0.13935 & 0.342146 & -0.23261 & 0.037976  & 0.542461 \\ \hline
$\overline{\delta}_{T_hOx}$       & -0.54236 & 0.202016 & 0.02244  & -0.0108  & 0.055535 & -0.16201 & 0.207578  & 0.047559 \\ \hline
$\overline{\delta}_{T_hIL_{10}}$  & -0.50803 & 0.108964 & 0.081498 & -0.05953 & -0.00172 & -0.3687  & 0.418748  & 0.01582  \\ \hline
$\overline{\delta}_{T_NOx}$       & -0.49523 & 0.074563 & 0.3709   & -0.1624  & -0.1892  & -0.08811 & -0.08204  & 0.005163 \\ \hline
$\overline{\delta}_{DOx}$         & 0.490666 & 0.083281 & -0.3979  & 0.293512 & 0.101234 & -0.01451 & 0.051204  & 0.086248 \\ \hline
$\overline{\lambda}_{MIL_{10}}$   & 0.460439 & -0.16563 & -0.02533 & -0.01806 & -0.1162  & 0.155653 & -0.16193  & -0.28758 \\ \hline
${\delta}_N$             & -0.17898 & -0.41799 & 0.060024 & 0.457777 & 0.195404 & -0.33921 & -0.2121   & -0.12769 \\ \hline
$\overline{\lambda}_{T_hIL_{12}}$ & 0.446693 & -0.14165 & 0.039065 & 0.002818 & -0.0831  & 0.271068 & -0.37524  & -0.02675 \\ \hline
$\overline{\lambda}_{IL_{6}A}$    & 0.041867 & 0.349583 & -0.13093 & 0.120269 & -0.27966 & 0.190116 & -0.0328   & -0.44345 \\ \hline
$\overline{\lambda}_{HD}$         & 0.131121 & 0.224023 & -0.0259  & -0.37941 & -0.02963 & 0.075347 & 0.020264  & 0.039692 \\ \hline
$\overline{\lambda}_{T_CD}$       & -0.21079 & 0.020634 & 0.342663 & -0.16486 & -0.20265 & -0.14179 & -0.04912  & 0.041677 \\ \hline
${\delta}_{T_r}$         & 0.021463 & -0.015   & -0.03453 & -0.03975 & 0.058484 & 0.227237 & -0.33124  & 0.095106 \\ \hline
$\overline{\lambda}_{CA}$         & 0.27029  & 0.317034 & -0.19047 & -0.19216 & -0.09968 & 0.134421 & 0.158812  & 0.046179 \\ \hline
$\overline{\lambda}_{IL_{10}T_h}$ & 0.108445 & -0.0392  & 0.047128 & 0.005142 & -0.06982 & -0.11446 & 0.262668  & -0.24719 \\ \hline
${\delta}_{T_h}$         & -0.2014  & 0.050374 & 0.025212 & -0.01961 & 0.00738  & -0.11924 & 0.139073  & 0.008914 \\ \hline
${\delta}_{E_N}$         & -0.2009  & 0.009339 & -0.00111 & -0.01816 & 0.055099 & 0.006399 & 0.004615  & -0.00835 \\ \hline
$\overline{\lambda}_{T_CIL_{12}}$ & -0.12113 & 0.011852 & 0.196798 & -0.09449 & -0.11657 & -0.08057 & -0.02713  & 0.023787 \\ \hline
$\overline{\delta}_{T_hT_r}$      & -0.193   & 0.035615 & 0.039436 & -0.02603 & -0.01069 & -0.1724  & 0.190377  & 0.004454 \\ \hline
$\overline{\lambda}_{T_hH}$       & 0.187532 & -0.04953 & 0.017159 & 0.008241 & -0.03951 & 0.111671 & -0.15939  & -0.00603 \\ \hline
$\overline{\lambda}_{IL_{10}T_r}$ & 0.055009 & -0.01372 & 0.031081 & 0.007731 & -0.04491 & -0.07817 & 0.18438   & -0.16578 \\ \hline
${\delta}_{M_N}$         & -0.18011 & -0.18367 & -0.14105 & -0.03822 & -0.09022 & 0.013848 & -0.01364  & -0.02507 \\ \hline
$\overline{\lambda}_{IL_{10}D}$   & 0.07282  & -0.02861 & 0.030803 & 0.001495 & -0.04589 & -0.07375 & 0.175919  & -0.16776 \\ \hline
$\overline{\delta}_{T_rOx}$       & 0.033514 & -0.00662 & -0.02321 & -0.01338 & 0.022394 & 0.116108 & -0.15454  & 0.048476 \\ \hline
$\overline{\delta}_{1}$           & -0.15452 & 0.07649  & -0.02509 & 0.090055 & -0.00688 & -0.07441 & -0.00521  & 0.085586 \\ \hline
$\overline{\lambda}_{VA}$         & 0.144249 & 0.029885 & 0.013258 & 0.039383 & -0.04812 & -0.01305 & -0.01899  & 0.030036 \\ \hline
$\overline{\lambda}_{HT_C}$       & 0.038301 & 0.077542 & -0.00686 & -0.12714 & -0.00907 & 0.031725 & 0.005395  & 0.017073 \\ \hline
$\overline{\lambda}_{CIL_6}$      & 0.110772 & 0.118184 & -0.0733  & -0.07816 & -0.03747 & 0.04971  & 0.060345  & 0.015981 \\ \hline
${\delta}_{D_N}$         & 0.101464 & 0.035046 & -0.08395 & -0.02353 & 0.02898  & -0.03294 & -0.01216  & -0.00558 \\ \hline
$\overline{\delta}_{COx}$         & -0.09977 & -0.07579 & 0.065286 & 0.015326 & 0.025504 & -0.00373 & -0.03561  & 0.000201 \\ \hline
$\overline{\delta}_{6}$           & -0.09955 & 0.026618 & -0.00451 & 0.016241 & 0.011826 & -0.0219  & -0.00156  & 0.021407 \\ \hline
$\overline{\delta}_{2}$           & -0.09194 & 0.054624 & -0.00974 & 0.050158 & -0.00497 & -0.03416 & -0.00535  & 0.044352 \\ \hline
$\overline{\lambda}_{VCOx}$       & 0.089004 & 0.016084 & 0.007842 & 0.022778 & -0.02888 & -0.00767 & -0.0111   & 0.017376 \\ \hline
$\overline{\lambda}_{IL_{12}M}$   & -0.08158 & -0.00314 & 0.011528 & -0.01296 & 0.018189 & 0.0137   & -0.01624  & -0.00635 \\ \hline
$\overline{\delta}_{T_COx}$       & -0.06644 & -0.02076 & -0.03514 & 0.028859 & 0.029179 & 0.03777  & -0.01164  & -0.03434 \\ \hline
$\overline{\lambda}_{DC}$         & 0.015601 & 0.03546  & -0.01632 & -0.06374 & 0.004352 & -0.03065 & -0.02127  & -0.0148  \\ \hline
${\delta}_{T_N}$         & -0.05709 & -0.00018 & 0.059686 & -0.02765 & -0.03341 & -0.01979 & -0.00934  & -0.00034 \\ \hline
$\overline{\delta}_{CT_C}$        & -0.05842 & -0.03021 & 0.040439 & 0.016297 & 0.002524 & -0.00239 & -0.01947  & 0.002348 \\ \hline
$\overline{\lambda}_{VOx}$        & 0.056838 & 0.005717 & 0.003987 & 0.011746 & -0.01738 & -0.00388 & -0.00601  & 0.008431 \\ \hline
$\overline{\lambda}_{IL_{10}M}$   & 0.025552 & -0.0104  & 0.009579 & 0.000308 & -0.01462 & -0.02253 & 0.052373  & -0.0508  \\ \hline
$\overline{\lambda}_{MIL_{12}}$   & 0.050651 & -0.01982 & -0.00266 & -0.0029  & -0.01242 & 0.016145 & -0.01659  & -0.03095 \\ \hline
$\overline{\delta}_{E_NOx}$       & -0.04738 & 0.001215 & 0.000104 & -0.00609 & 0.014329 & 0.00168  & 0.00186   & -0.00229 \\ \hline
$\overline{\lambda}_{IL_{10}C}$   & 0.024281 & -0.00891 & 0.009008 & 0.001184 & -0.01372 & -0.02127 & 0.046414  & -0.04441 \\ \hline
$\overline{\lambda}_{DOx}$        & 0.009237 & 0.021903 & -0.00997 & -0.04082 & 0.003702 & -0.01876 & -0.01331  & -0.0095  \\ \hline
$\overline{\delta}_{3}$           & -0.04044 & 0.020099 & -0.0024  & 0.014874 & 0.000851 & -0.01059 & -0.00192  & 0.014028 \\ \hline
$\overline{\lambda}_{IL_{6}Ox}$   & 0.00573  & 0.031329 & -0.01187 & 0.009011 & -0.02495 & 0.016517 & -0.00235  & -0.03829 \\ \hline
$\overline{\lambda}_{IL_{6}MOx}$  & 0.004933 & 0.027207 & -0.01033 & 0.007862 & -0.02162 & 0.014438 & -0.00206  & -0.03347 \\ \hline
$\overline{\delta}_{4}$           & -0.0276  & 0.009627 & -0.00122 & 0.006177 & 0.002484 & -0.00616 & -0.00068  & 0.006957 \\ \hline
${\delta}_{Ox}$          & -0.02736 & 0.014083 & -0.00205 & 0.011488 & -7.9E-05 & -0.00827 & -0.00137  & 0.010533 \\ \hline
$\overline{\lambda}_{IL_{10}T_C}$ & 0.007177 & -0.00161 & 0.004273 & 0.001135 & -0.00599 & -0.01061 & 0.025     & -0.02238 \\ \hline
$\overline{\lambda}_{T_rOx}$      & -0.02465 & 7.32E-05 & 0.024956 & -0.0097  & -0.01463 & -0.01503 & 0.007547  & -0.00354 \\ \hline
$\overline{\delta}_{D_NOx}$       & 0.021962 & 0.010357 & -0.01643 & -0.00565 & 0.003955 & -0.00754 & -0.00248  & -0.0009  \\ \hline
$\overline{\lambda}_{IL_{10}Ox}$  & 0.011223 & -0.00474 & 0.004007 & 4.72E-07 & -0.00578 & -0.00881 & 0.020689  & -0.02038 \\ \hline
$\overline{\lambda}_{IL_{6}M}$    & 0.004059 & 0.016726 & -0.0066  & 0.004112 & -0.01323 & 0.008869 & -0.00108  & -0.02056 \\ \hline
$\overline{\delta}_{CT_COx}$      & -0.01937 & -0.01234 & 0.015427 & 0.004354 & 0.000709 & -0.00087 & -0.00728  & 0.000627 \\ \hline
$\overline{\delta}_{T_CT_r}$      & -0.01768 & -0.00685 & -0.00922 & 0.004857 & 0.003968 & 0.00675  & 0.005594  & -0.0132  \\ \hline
${\delta}_{T_C}$         & -0.01265 & -0.00471 & -0.00634 & 0.003719 & 0.004278 & 0.005555 & 0.001281  & -0.00813 \\ \hline
$\overline{\lambda}_{IL_{6}D}$    & 0.002391 & 0.008172 & -0.00331 & 0.001679 & -0.00642 & 0.004329 & -0.00044  & -0.01005 \\ \hline
$\overline{\delta}_{T_CIL_{10}}$  & -0.00835 & -0.00309 & -0.00417 & 0.002305 & 0.002485 & 0.00344  & 0.00143   & -0.00558 \\ \hline
$Ox_{crit}^{(1)}$      & 0.006543 & -0.0013  & 0.000865 & -0.0016  & -0.00205 & -0.0018  & -7.64E-05 & -0.00278 \\ \hline
\caption{Full principle component sensitivity results for Mouse 2.}
\end{longtable}

\vspace{0.6in}
\begin{longtable}{|l|l|l|l|l|l|l|l|l|}
\hline
Parameters             & pc 1      & pc 2      & pc 3      & pc 4      & pc 5      & pc 6      & pc 7      & pc 8      \\ \hline
\endfirsthead
\hline
Parameters             & pc 1      & pc 2      & pc 3      & pc 4      & pc 5      & pc 6      & pc 7      & pc 8      \\ 
\endhead 
\endfoot
\hline
$\overline{A}_{D_N}$              & -0.87505 & -2.94752  & -0.69943  & -0.16148  & 0.464956  & -0.15158  & -0.32059  & 0.162883  \\ \hline
$\overline{\lambda}_{VE_N}$       & 2.58038  & -0.25886  & -0.25373  & 0.315519  & 0.453465  & 1.039879  & 0.416721  & 0.016513  \\ \hline
$Ox_{crit}^{(2)}$      & -2.5587  & -0.00099  & 0.369021  & 0.272826  & 0.016675  & 0.624977  & 0.023624  & 0.298538  \\ \hline
$\overline{\lambda}_{OxE_N}$      & 2.541831 & 0.005742  & -0.37717  & -0.27359  & -0.00885  & -0.63703  & -0.02699  & -0.28755  \\ \hline
$\overline{\delta}_{5}$           & -2.19643 & -0.00166  & 0.33925   & 0.216156  & -0.01237  & 0.529533  & 0.017523  & 0.242622  \\ \hline
$\overline{A}_M$                  & -1.12537 & 0.606416  & -1.99833  & -0.0746   & -0.61427  & 0.091433  & 0.076939  & -0.02143  \\ \hline
${\delta}_{IL_{12}}$     & 1.941517 & 0.207379  & -0.60539  & 0.037548  & 0.586014  & -0.04191  & 0.325014  & 0.461415  \\ \hline
$\overline{\lambda}_A$            & -1.80728 & 0.508225  & 0.173992  & -1.45238  & 0.121842  & -0.19281  & -0.19865  & 0.07338   \\ \hline
${\delta}_{D}$           & 0.861871 & 1.739064  & 0.068417  & 0.171211  & 0.293366  & 0.074351  & -0.55233  & 0.230786  \\ \hline
$\overline{\lambda}_{T_CIL_{12}}$ & -1.55047 & 0.177748  & -0.16592  & 0.112431  & 0.237274  & -0.14751  & 0.383719  & -0.2082   \\ \hline
$\overline{A}_{T_N}$              & 1.385577 & -0.18858  & 0.27303   & -0.32866  & -0.56372  & -0.54376  & 0.444367  & 0.845971  \\ \hline
$\overline{\delta}_{E_NIL_{12}}$  & -1.35533 & 0.13154   & 0.132268  & -0.1669   & -0.2334   & -0.5324   & -0.20818  & -0.00358  \\ \hline
${\delta}_{M}$           & 0.111715 & -0.14361  & 1.21105   & 0.14577   & 1.094694  & -0.49165  & 0.205021  & -0.12885  \\ \hline
${\delta}_V$             & -0.89404 & -0.02171  & 0.101052  & 1.205259  & -0.44265  & -0.59492  & -0.23083  & 0.079252  \\ \hline
$\overline{\delta}_{T_hIL_{10}}$  & -1.13773 & 0.374212  & -0.4449   & 0.167406  & 0.592332  & -0.19184  & 0.237802  & -0.06636  \\ \hline
$\overline{\delta}_{T_CT_r}$      & 1.134284 & -0.09582  & 0.04201   & -0.09484  & -0.14477  & 0.106838  & -0.38688  & -0.00204  \\ \hline
$\overline{\lambda}_{VA}$         & 0.318603 & 0.075819  & -0.04503  & -1.1334   & 0.2958    & 0.29177   & 0.11182   & -0.08413  \\ \hline
${\delta}_A$             & 1.002255 & -0.20518  & -0.1329   & 0.674658  & -0.14503  & -0.22512  & 0.034027  & -0.08288  \\ \hline
${\delta}_{IL_{10}}$     & 0.992387 & -0.35396  & 0.381568  & -0.19339  & -0.63234  & 0.093846  & -0.09687  & -0.02125  \\ \hline
$\overline{\lambda}_{IL_{10}T_C}$ & -0.9691  & 0.346442  & -0.37412  & 0.18841   & 0.618861  & -0.09245  & 0.092751  & 0.020632  \\ \hline
$\overline{\lambda}_{MT_h}$       & 0.890217 & -0.36908  & 0.533368  & -0.07028  & -0.42776  & 0.329935  & -0.19334  & 0.10402   \\ \hline
$\overline{\lambda}_{T_hIL_{12}}$ & 0.842242 & -0.32804  & 0.361679  & -0.10643  & -0.46864  & 0.326561  & -0.42393  & -0.19587  \\ \hline
$\overline{\lambda}_{T_rOx}$      & 0.795008 & 0.009964  & -0.207    & 0.008974  & 0.171963  & -0.00924  & -0.23656  & -0.47251  \\ \hline
$\overline{\delta}_{DC}$          & 0.333491 & 0.7514    & 0.050984  & 0.05185   & 0.086142  & -0.02111  & -0.21594  & 0.0927    \\ \hline
$\overline{\lambda}_{DOx}$        & -0.35    & -0.0941   & 0.314482  & -0.05524  & -0.53486  & 0.074853  & 0.738862  & -0.33326  \\ \hline
$\overline{\delta}_{T_rOx}$       & -0.72276 & 0.024305  & 0.089153  & 0.056877  & 0.004469  & 0.122136  & 0.009916  & 0.072773  \\ \hline
$\overline{\lambda}_{IL_{12}D}$   & -0.56482 & -0.14522  & 0.205327  & 0.074905  & -0.16399  & 0.281632  & -0.07769  & -0.22667  \\ \hline
${\delta}_{T_r}$         & -0.4568  & -0.00755  & 0.056072  & 0.075045  & 0.02658   & 0.170864  & 0.064217  & 0.061528  \\ \hline
${\delta}_H$             & 0.181997 & -0.02301  & -0.20772  & 0.093154  & 0.355488  & 0.0829    & -0.43593  & 0.09741   \\ \hline
$\overline{\lambda}_{HT_C}$       & -0.17798 & 0.023467  & 0.203621  & -0.09163  & -0.34843  & -0.08162  & 0.427893  & -0.09509  \\ \hline
${\delta}_{D_N}$         & 0.097228 & 0.359805  & 0.092065  & 0.018873  & -0.06692  & 0.015914  & 0.046809  & -0.02417  \\ \hline
$\overline{\lambda}_{DH}$         & -0.13037 & 0.036455  & 0.144707  & -0.05291  & -0.24388  & -0.03311  & 0.334583  & -0.12813  \\ \hline
$\overline{\delta}_{6}$           & -0.30386 & -0.00503  & 0.050572  & 0.039086  & 0.003679  & 0.099323  & 0.008127  & 0.038139  \\ \hline
$\overline{\lambda}_C$            & 0.15685  & 0.256998  & -0.02401  & -0.00589  & 0.100288  & 0.07134   & -0.10218  & 0.019473  \\ \hline
${\delta}_C$             & -0.15666 & -0.24425  & 0.028112  & -0.01015  & -0.08918  & -0.03017  & 0.061826  & 0.010763  \\ \hline
$\overline{\delta}_{T_CIL_{10}}$  & 0.217173 & -0.02892  & 0.047722  & 0.034712  & 0.031527  & 0.153054  & -0.17147  & 0.014327  \\ \hline
$\overline{\lambda}_{CA}$         & 0.106002 & 0.188766  & -0.01418  & -0.01383  & 0.058318  & 0.021267  & -0.04975  & 0.007865  \\ \hline
$\overline{\lambda}_{MIL_{12}}$   & 0.138511 & -0.08617  & 0.188353  & -0.00482  & -0.08195  & 0.085574  & -0.07187  & 0.043235  \\ \hline
${\delta}_{IL_6}$        & -0.02989 & -0.0567   & 0.181174  & 0.027823  & 0.151255  & -0.08758  & 0.048638  & -0.05565  \\ \hline
${\delta}_{E_N}$         & -0.17593 & 0.007289  & 0.02277   & 0.004323  & -0.01764  & -0.01469  & -0.01652  & 0.007006  \\ \hline
$\overline{\lambda}_{T_CD}$       & -0.14362 & 0.011179  & -0.00545  & 0.021037  & 0.029502  & 0.012399  & 0.027115  & -0.01655  \\ \hline
$\overline{\lambda}_{IL_{12}M}$   & -0.12721 & -0.03871  & 0.047449  & 0.020866  & -0.03656  & 0.080907  & -0.01669  & -0.05828  \\ \hline
$\overline{\lambda}_{IL_{6}M}$    & 0.014065 & 0.03246   & -0.12266  & -0.01924  & -0.10425  & 0.053338  & -0.02971  & 0.03543   \\ \hline
$\overline{\delta}_{CT_C}$        & -0.08213 & -0.11819  & 0.016707  & -0.00831  & -0.05328  & -0.03512  & 0.041751  & -0.00108  \\ \hline
$\overline{\delta}_{CT_COx}$      & -0.07342 & -0.10512  & 0.016484  & -0.00511  & -0.04616  & -0.0256   & 0.025333  & 0.002063  \\ \hline
$\overline{\lambda}_{CIL_6}$      & 0.045753 & 0.073257  & -0.00716  & -0.00052  & 0.030325  & 0.023894  & -0.03242  & 0.006405  \\ \hline
$\overline{\delta}_{T_COx}$       & 0.070864 & -0.01122  & 0.003388  & 0.011485  & 0.00374   & 0.035452  & -0.00701  & 0.002625  \\ \hline
$\overline{\alpha}_{NC}$          & -0.03976 & 0.000548  & 0.024595  & -0.04384  & -0.02358  & 0.009124  & -0.04275  & 0.0435    \\ \hline
$Ox_{crit}^{(1)}$      & 0.028058 & -0.00645  & 0.023499  & -0.0017   & -0.0164   & 0.026118  & -0.00338  & -0.01213  \\ \hline
$\overline{\lambda}_{VCOx}$       & 0.007165 & 0.001441  & -0.00099  & -0.02278  & 0.006119  & 0.006239  & 0.002419  & -0.00167  \\ \hline
${\delta}_{M_N}$         & 0.012753 & -0.00706  & 0.022506  & 0.001262  & 0.006739  & -8.25E-05 & -0.00068  & 0.000461  \\ \hline
$\overline{\delta}_{1}$           & -0.01972 & -0.00017  & 0.003995  & 0.001148  & -0.00086  & 0.005215  & 0.00014   & 0.002313  \\ \hline
$\overline{\delta}_{3}$           & -0.01616 & 0.000542  & 0.003018  & -0.00055  & -0.00171  & 0.0006    & -0.00058  & 0.001154  \\ \hline
$\overline{\delta}_{COx}$         & -0.00955 & -0.01355  & 0.00231   & -0.00051  & -0.00565  & -0.00212  & 0.001642  & 0.000824  \\ \hline
$\overline{\lambda}_{T_hD}$       & 0.011997 & -0.00424  & 0.003611  & -0.00213  & -0.00699  & 0.003067  & -0.00491  & -0.00345  \\ \hline
$\overline{\delta}_{T_hT_r}$      & -0.01002 & 0.001867  & -0.00141  & 0.003511  & 0.006268  & 0.002775  & 0.003185  & 0.000753  \\ \hline
$\overline{\delta}_{4}$           & -0.00969 & -2.10E-05 & 0.00188   & 0.000664  & -0.00039  & 0.002387  & -5.11E-06 & 0.001056  \\ \hline
$\overline{\lambda}_{DC}$         & -0.00424 & -0.0013   & 0.003773  & -0.00043  & -0.00637  & 0.001495  & 0.009071  & -0.00411  \\ \hline
${\delta}_{T_h}$         & -0.00544 & 0.001324  & -0.00123  & 0.001398  & 0.003063  & 0.000434  & 0.001356  & 0.00012   \\ \hline
${\delta}_{T_C}$         & 0.005205 & -0.00063  & 0.000582  & 0.000332  & 0.000126  & 0.002221  & -0.00244  & 0.000179  \\ \hline
$\overline{\lambda}_{T_hH}$       & 0.004562 & -0.00184  & 0.001934  & -0.00035  & -0.00235  & 0.002078  & -0.00183  & -0.00094  \\ \hline
$\overline{\lambda}_{MIL_{10}}$   & 0.003302 & -0.00199  & 0.003968  & 5.28E-05  & -0.00184  & 0.002192  & -0.00136  & 0.000929  \\ \hline
${\delta}_N$             & 0.003355 & 6.09E-06  & -0.00207  & 0.00358   & 0.002003  & -0.00086  & 0.003484  & -0.00371  \\ \hline
$\overline{\lambda}_{VOx}$        & 0.001748 & 0.000178  & -0.00022  & -0.0037   & 0.001102  & 0.001233  & 0.000494  & -0.00025  \\ \hline
$\overline{\lambda}_{IL_{10}T_r}$ & -0.00328 & 0.000589  & -0.00042  & 0.001187  & 0.002086  & 0.001009  & 0.001011  & 0.00037   \\ \hline
$\overline{\lambda}_{VM}$         & 0.001767 & 0.000115  & -0.00021  & -0.00311  & 0.001011  & 0.001238  & 0.000483  & -0.00021  \\ \hline
$\overline{\lambda}_{IL_{10}C}$   & -0.00295 & 0.00063   & -0.00048  & 0.000895  & 0.001741  & 0.000547  & 0.0007    & 0.000257  \\ \hline
$\overline{\lambda}_{IL_{6}MOx}$  & 0.000259 & 0.000649  & -0.00268  & -0.00042  & -0.0023   & 0.00112   & -0.00061  & 0.000756  \\ \hline
$\overline{\lambda}_{IL_{6}D}$    & 0.000357 & 0.000733  & -0.00262  & -0.0004   & -0.00221  & 0.001195  & -0.00065  & 0.000776  \\ \hline
${\delta}_{T_N}$         & -0.00252 & 0.00017   & -0.00027  & 0.000688  & 0.00093   & 0.001276  & -0.00068  & -0.00153  \\ \hline
$\overline{\delta}_{2}$           & -0.00227 & 5.85E-05  & 0.00045   & -7.41E-05 & -0.00024  & 0.000159  & -7.04E-05 & 0.000182  \\ \hline
$\overline{\lambda}_{IL_{6}A}$    & 0.000154 & 0.00047   & -0.0021   & -0.00033  & -0.00181  & 0.00083   & -0.00046  & 0.000575  \\ \hline
$\overline{\lambda}_{IL_{10}Ox}$  & -0.00203 & 0.000486  & -0.0004   & 0.00056   & 0.001194  & 0.000253  & 0.000454  & 0.000149  \\ \hline
$\overline{\lambda}_{IL_{10}T_h}$ & -0.00131 & 0.000257  & -0.0002   & 0.000451  & 0.000839  & 0.000356  & 0.00038   & 0.00014   \\ \hline
$\overline{\lambda}_{HC}$         & -0.00074 & -0.00023  & 0.000668  & -0.00013  & -0.0011   & 0.000188  & 0.001299  & -0.00042  \\ \hline
$\overline{\delta}_{E_NOx}$       & -0.00125 & 6.13E-05  & 0.000168  & 5.80E-05  & -0.00013  & -0.00012  & -0.00015  & 3.33E-05  \\ \hline
$\overline{\lambda}_{IL_{10}M}$   & -0.00115 & 0.000265  & -0.00022  & 0.000334  & 0.000685  & 0.000175  & 0.00026   & 8.98E-05  \\ \hline
$\overline{\lambda}_{IL_{10}D}$   & -0.001   & 0.000255  & -0.00022  & 0.000262  & 0.000594  & 9.24E-05  & 0.000202  & 6.55E-05  \\ \hline
$\overline{\delta}_{D_NOx}$       & 0.000184 & 0.000638  & 0.000153  & 1.63E-05  & -0.00011  & 4.20E-06  & 9.25E-05  & -4.00E-05 \\ \hline
$\overline{\lambda}_{T_rD}$       & 0.000613 & 2.23E-06  & -0.00013  & 3.04E-05  & 0.000147  & 7.28E-05  & -0.00023  & -0.00031  \\ \hline
$\overline{\delta}_{DOx}$         & 0.000307 & 0.000573  & 5.43E-06  & 5.47E-05  & 0.000111  & 3.84E-05  & -0.00014  & 7.36E-05  \\ \hline
${\delta}_{Ox}$          & -0.00056 & 1.53E-05  & 0.000121  & -2.49E-05 & -6.84E-05 & 3.86E-05  & -2.18E-05 & 4.36E-05  \\ \hline
$\overline{\lambda}_{HM}$         & -0.00027 & -7.01E-05 & 0.000255  & -6.15E-05 & -0.00042  & 4.00E-05  & 0.000503  & -0.00015  \\ \hline
$\overline{\lambda}_{HD}$         & -0.00025 & -4.97E-05 & 0.000241  & -7.28E-05 & -0.0004   & 1.04E-06  & 0.000463  & -0.00014  \\ \hline
$\overline{\delta}_{T_hOx}$       & -0.00039 & 0.000115  & -0.00011  & 5.77E-05  & 0.000193  & -4.50E-05 & 6.17E-05  & -5.43E-06 \\ \hline
$\overline{\lambda}_{HN}$         & -0.00018 & -7.24E-05 & 0.000162  & -2.82E-05 & -0.00027  & 6.28E-05  & 0.000348  & -0.00011  \\ \hline
$\overline{\lambda}_{IL_{6}Ox}$   & 3.10E-05 & 7.19E-05  & -0.00029  & -4.49E-05 & -0.00025  & 0.000123  & -6.63E-05 & 8.22E-05  \\ \hline
$\overline{\delta}_{T_NOx}$       & -0.00011 & 1.49E-05  & -1.59E-05 & 2.08E-05  & 3.56E-05  & 2.91E-05  & -2.72E-05 & -5.84E-05 \\ \hline
\caption{Full principle component sensitivity results for Mouse 3.}
\end{longtable}

\begin{landscape}
  \begin{figure}
    \centering
    \includegraphics[scale=0.75]{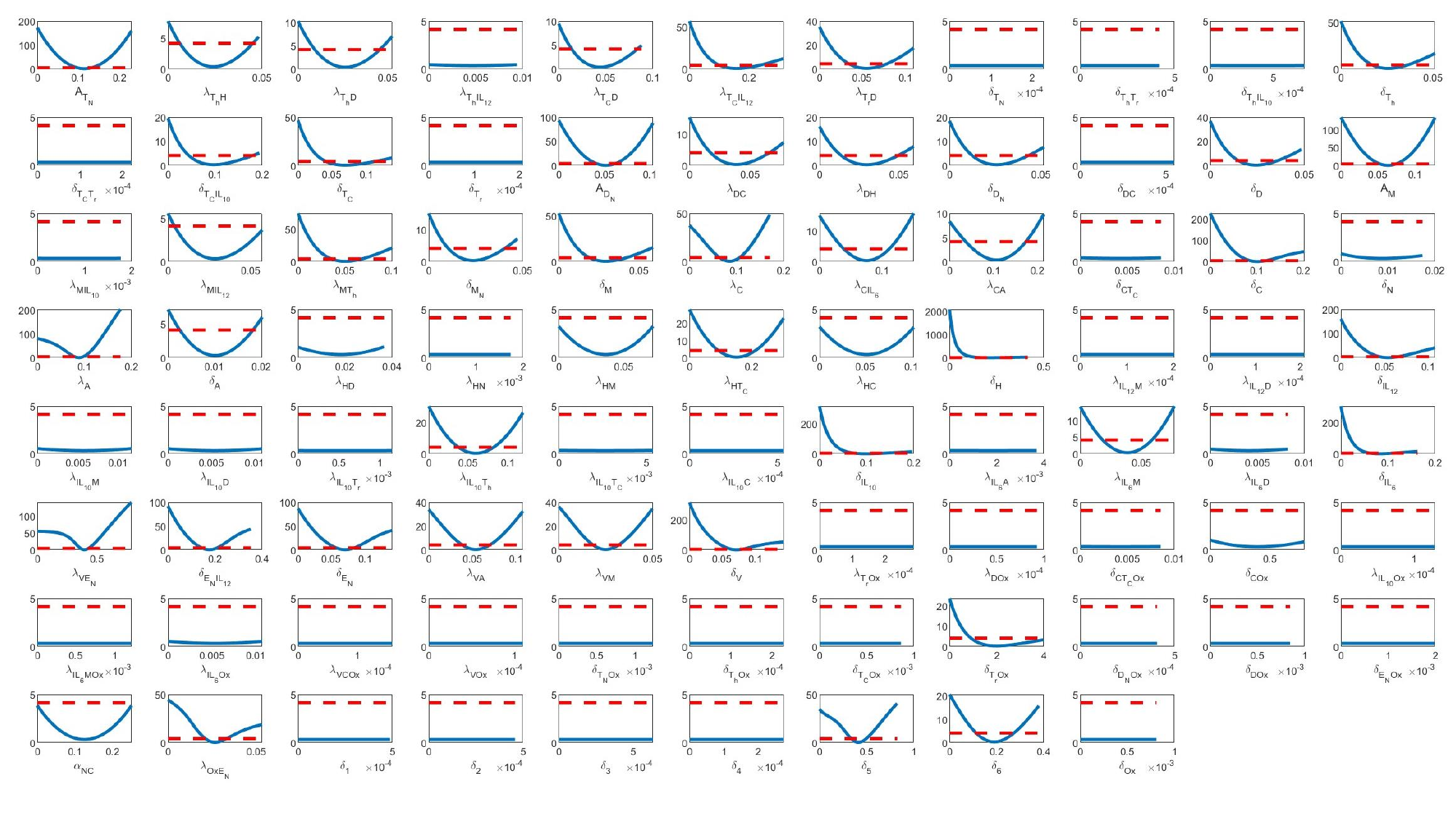}
    \caption{Parameters profile likelihood plots for Mouse 1. The y-axis is the value of the profile likelihood. The dashed line is the threshold $\Delta_{\alpha}$ for $\alpha=0.95$.}
    \label{fig:landscape}
  \end{figure}
\end{landscape}

\begin{landscape}
  \begin{figure}
    \centering
    \includegraphics[scale=0.75]{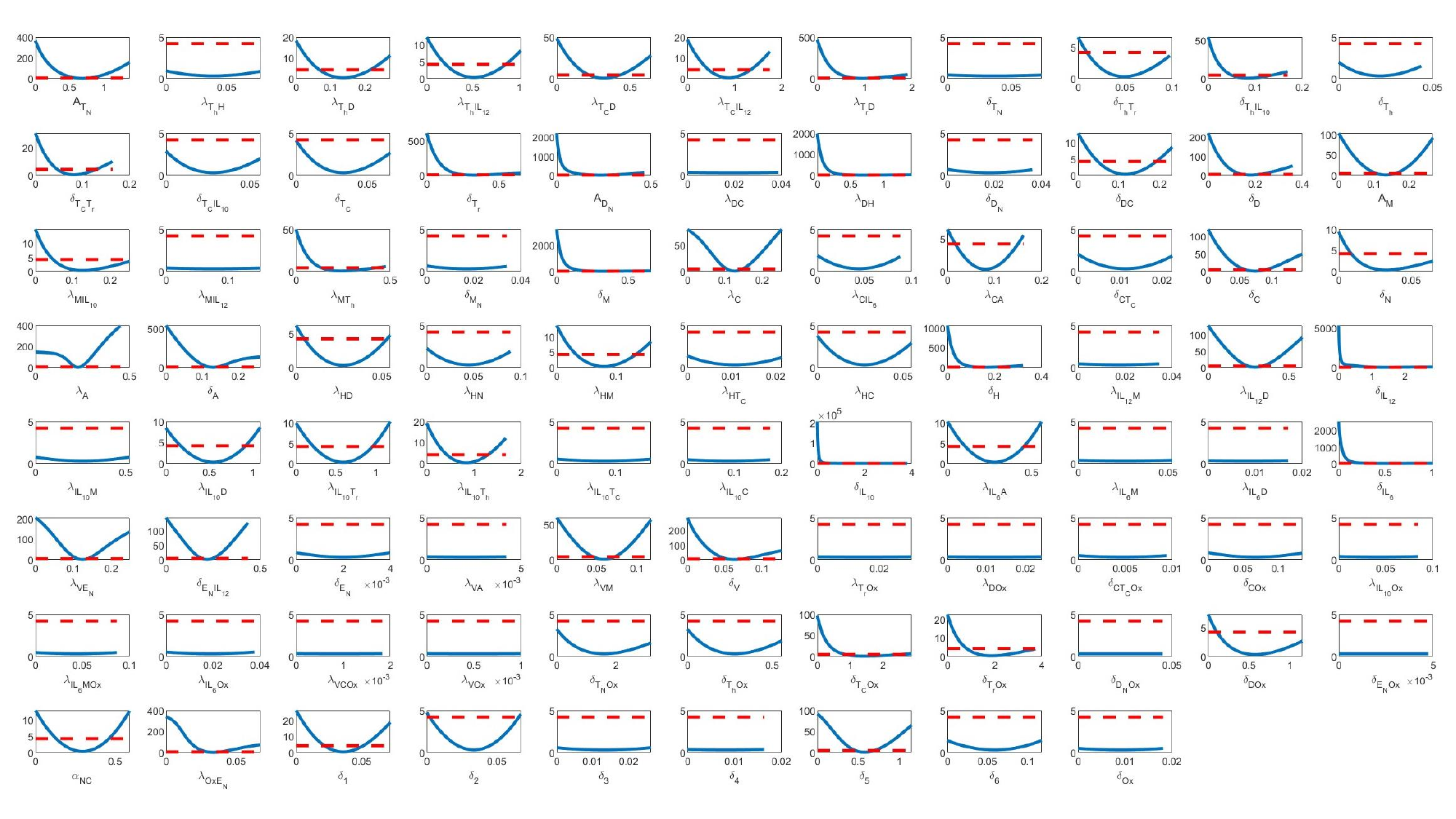}
    \caption{Parameters profile likelihood plots for Mouse 2. The y-axis is the value of the profile likelihood. The dashed line is the threshold $\Delta_{\alpha}$ for $\alpha=0.95$.}
    \label{fig:landscape}
  \end{figure}
\end{landscape}

\begin{landscape}
  \begin{figure}
    \centering
    \includegraphics[scale=0.75]{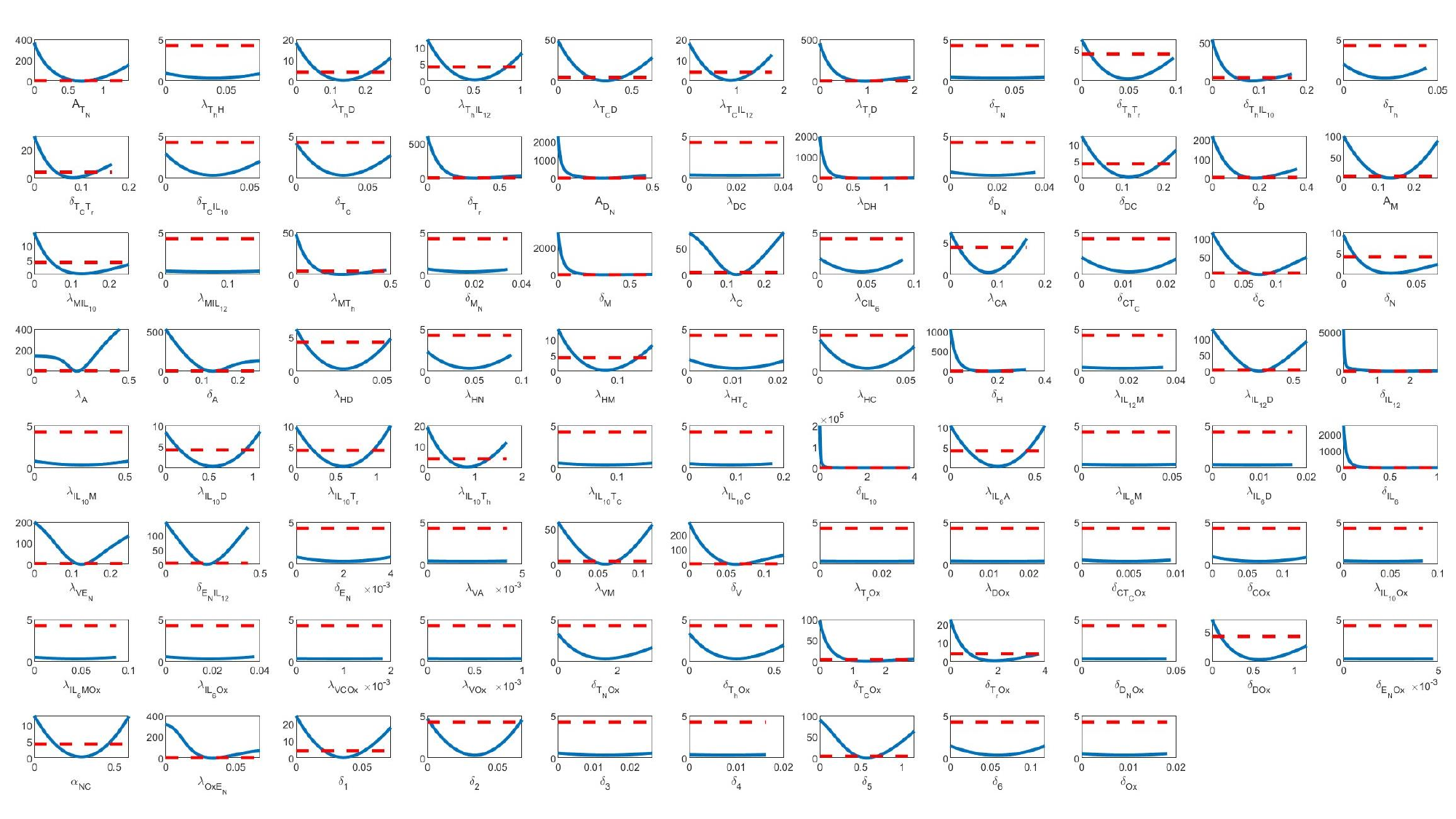}
    \caption{Parameters profile likelihood plots for Mouse 3. The y-axis is the value of the profile likelihood. The dashed line is the threshold $\Delta_{\alpha}$ for $\alpha=0.95$.}
    \label{fig:landscape}
  \end{figure}
\end{landscape}

% keywords can be removed

  %%% Uncomment this line and comment out the ``thebibliography'' section below to use the external .bib file (using bibtex) .





% \end{thebibliography}